\documentclass[12pt]{article}

\usepackage{cite}
\usepackage{times}
\usepackage{float,wrapfig,epsfig,alltt,amsmath,amstext,amssymb,tabularx,fanc
ybox,picinpar}
\usepackage{graphicx}
\usepackage{subfigure}
\usepackage{lscape,color}
\usepackage{float,epsfig,amsmath,amstext,amssymb}
\usepackage{epic,eepic,tabularx}
\usepackage{subfigure}
\usepackage{url}
\begin{document}

\newtheorem{lemma}{Lemma}

\newtheorem{theorem}[lemma]{Theorem}

\newtheorem{fact}[lemma]{Fact}

\newtheorem{corollary}[lemma]{Corollary}

\newtheorem{observation}[lemma]{Observation}

\newtheorem{definition}[lemma]{Definition}

\newtheorem{proposition}[lemma]{Proposition}

%\title{ Constructing the Feasible Regions for Placement of Geometric Objects 
%with Costraints on Their Relative Position: the Case of Translational 
%Geometric Situations}
%%%\title{ Analysis of relative position and constructing the set of feasible 
%%%positions of geometric objects: the case of the translational geometric 
%%%situations}
%\title{ Analysis of relative position and synthesis of feasible 
%positions of geometric objects: the case of the translational geometric 
%situations}
\title{ Spatial planning with constraints on translational distances between 
geometric objects 
\thanks{ A full version of this paper is available at \cite{Pk1}.}
%\thanks{ A full version of this paper is available at 
%{\tt http://www.cs.tau.ac.il/\verb1~1genap/SpatPlan.ps}, 
%{\tt http://www.cs.tau.ac.il/\verb1~1genap/SpatPlan.pdf}.
%} 
}

%\thanks
%{  ****************************
%}
%\author{}
\author{Gennady Pustylnik \thanks{ PG--Consulting, Holon 58371, Israel. 
{\it Email addreses:} {\tt genap@post.tau.ac.il}, {\tt gennadypus@walla.com}.}
}
%\date{ }

\maketitle

\begin{abstract}
The main constraint on relative position of geometric objects, used in spatial 
planning for computing the C-space maps (for example, in robotics, CAD, and 
packaging), is the relative non-overlapping of objects. This is the simplest 
constraint in which the minimum translational distance between objects is 
greater than zero, or more generally, than some positive value. We present a 
technique, based on the Minkowski operations, for generating the translational 
C-space maps for spatial planning with more general and more complex 
constraints on the relative position of geometric objects, such as constraints 
on various types (not only on the minimum) of the translational distances 
between objects. The developed technique can also be used, respectively, for 
spatial planning with constraints on translational distances in a given 
direction, and rotational distances between geometric objects, as well as  for 
spatial planning with given dynamic geometric situation of moving objects.

%{\it Keywords:} Spatial planning; Configuration space; Minkowski operations
\end{abstract} 


%\newpage
%---------------------------------------------------------------------------

\def\bd{{\partial}}
\def\vbd{{\vec \partial}}
\def\bsl{{\backslash}}
\def\bsla{{\big\backslash}}
\def\bslb{{\Big\backslash}}
\def\a{{\alpha}}
\def\g{{\gamma}}
\def\G{{\Gamma}}
\def\T{{\Theta}}
\def\t{{\theta}}
\def\l{{\lambda}}
\def\u{{\tau}}
\def\e{{\varepsilon}}
\def\dl{{\delta}}
\def\Dl{{\Delta}}
\def\+{{\oplus}}
\def\-{{\ominus}}
\def\s{{\subset}}
\def\se{{\subseteq}}
\def\sp{{\supseteq}}
\def\A{{\cal A}}
\def\B{{\cal B}}
\def\C{{\cal C}}
\def\D{{\cal D}}
\def\F{{\cal F}}
\def\N{{\cal N}}
\def\S{{\cal S}}
\def\P{{\cal P}}
\def\R{{\cal R}}
\def\U{{\cal U}}
\def\es{{\emptyset}}
\def\vAi{{\vec A_i}}
\def\vai{{\vec a_i}}
\def\vBj{{\vec B_j}}
\def\vbj{{\vec b_j}}
\def\cB{{\check B}}
\def\cA{{\check A}}
\def\hA{{\hat A}}
\def\ha{{\hat a}}
\def\hB{{\hat B}}
\def\vA{{\vec A}}
\def\va{{\vec a}}
\def\o{{\omega}}
\def\O{{\Omega}}
\def\vp{{\varphi}}
\def\CT{{\cal T}}
\def\p2{{\frac{\pi}{2}}}
\def\i{{\mathbf{i}}}
\def\k{{\mathbf{k}}}
\def\ll{{\parallel}}
\def\sw{{\sl sweep}}
\def\us{{\sl unsweep}}
\def\hM{{\hat M}}
\def\bm{{B^A_{\max}}}
\def\am{{A^A_{\max}}}
\def\ab{{A\+ B}}
\def\abi{{\i A\+\i B}}
\def\adb{{A\+\check B}}
\def\adbi{{\i A\+\i\check B}}
\def\amb{{A\- B}}
\def\ambi{{\i A\-\i B}}
\def\aeb{{A\-\check B}}
\def\aebi{{\i A\-\i\check B}}
\def\bea{{\check B\- A}}
\def\beai{{\i\check B\- \i A}}
\def\fa{{\bf A}}
\def\vPi{{\vec P_i}}
\def\vP{{\vec P}}
\def\vQj{{\vec Q_j}}
\def\vQ{{\vec Q}}
%\newpage

%----------------------------1-------------------------
\subsection*{1 Introduction}

Problems concerning the relative placement of geometric objects are called 
{\it spatial planning} problems \cite{LP}. Such problems are important in 
robotics \cite{LT}, collision detection \cite{JTT}, \cite{LinM}, and 
computer-aided design and manufacturing (CAD/CAM) \cite{EK}, \cite{St1}.

A technique commonly used for solving spatial planning problems is the 
{\it configuration space} (or {\it C-space}) approach, based on representing 
each placement of an object, i.e., its position and orientation, as a point in 
some parametric C-space \cite{LPW}, \cite{LP}. (Each coordinate of the C-space 
represents a degree of freedom in the position or orientation of the object.) 

Given a collection of objects, the {\it translational} spatial planning 
problem consists in computing the set of all the feasible positions (the 
orientations are fixed) of the objects, where certain constraints on their 
relative position are specified. The feasible region of (placements of) an 
object is called the {\it free} C-space of the object. The prohibited 
configurations of an object form a {\it forbidden} region. The {\it C-space 
mapping} for a particular spatial planning problem consists in partitioning 
the C-space into free and forbidden regions, where the latter are called 
{\it C-space obstacles}. See \cite{LP} and \cite{WB} for more details.

\subsubsection*{1.1 Previous and related work}

Detailed surveys of previous work on spatial planning can be found in 
\cite{DM}, \cite{G}, \cite{JTT}, \cite{LinM}, \cite{LP}, and \cite{WB}. See 
also \cite{AB}, \cite{BKOS}, \cite{LM},  \cite{Ml1}, \cite{St1}, and 
\cite{SYv} and references therein for other related works on placement and 
spatial planning. 

The basic constraint on the relative position of geometric objects, used in 
spatial planning for generating the C-space maps, is the relative 
non-overlapping of objects; here the minimum translational distance between 
the objects must be greater than zero, or more generally, than some positive 
value. Indeed, this is a key requirement in robotics, packaging, and nesting. 
See, e.g., \cite{BKOS}, \cite{LM}, \cite{LinM},\cite{St}, and \cite{WB}. 
However, the geometric problems arising in design and manufacturing require a 
placement of objects with more complex constraints on their relative position, 
such as constraints on the minimum and/or maximum translational distances 
between objects, and/or their Hausdorff distances. The problem of generating 
the C-maps with such complex constraints is also interesting theoretically. 

Placement problems taking into account the minimum translational distance 
(MTD) between objects arise in industrial applications, concerning the cutting 
of materials, the layout of templates on a stock material, and the layout of 
an IC chip with geometric design constraints. See, e.g., \cite{DM}, \cite{LP}, 
\cite{Ml1}, \cite{St}, and \cite{St1}. In these  problems the MTD has to be at 
least the cutting tolerance of the machine that cuts the shapes out of stock 
material, or the minimal feasible distances between electronic modules of an 
IC chip. Placement problems with constraints on the MTD between objects have 
been formulated in \cite{St} and \cite{St1}. The papers \cite{P} and 
\cite{St1} have considered placement problems with consrtaints on the minimal 
value of MTD between objects, and placement problems with constraints on both 
the minimal and maximal admissible values of MTD have been considered in 
\cite{St1}. Algorithms for solving various placement problems with constraints 
on the MTD have been considered in \cite{P}, \cite{St1}, and 
\cite{SPV} -- \cite{SYv}.

The work in \cite{Pk} has solved the problem of placement of a pair of objects 
with constraints on the {\it minimum} and {\it maximum} translational 
distances between them, and has also considered distances involving 
containment of objects. Placement problems with constraints on several types 
of {\it translational} distances between objects, including directional 
Hausdorff distances between objects, have been studied in \cite{Pk2} and 
\cite{Pk3}. These works have also studied a Boolean function, called the 
{\it geometric situation}, which describes the system of constraints on the 
relative position of objects, and have formulated and solved placement 
problems with several other types of geometric situations, namely, with 
rotational and dynamic geometric situations.

An algorithm for computing the minimum Hausdorff distance between two planar 
objects under translation is given in \cite{AST}. (Efficient computation of 
Hausdorff distances has applications, e.g., in pattern recognition and 
computer vision; see \cite{AST} and \cite{Se1}.) This work has also studied 
placement problems taking into account bidirectional Hausdorff distances 
between objects. 

This paper presents a technique, based on the Minkowski operations, for 
generating the translational C-space maps for spatial planning problems with 
more general and more complex constraints on the relative position of 
geometric objects. This technique is an extention of that reported in 
\cite{Pk2} and \cite{Pk3}. The developed technique can also be used, 
respectively, for spatial planning with constraints on {\it translational 
distances in a given direction}, and/or on {\it rotational} distances between 
geometric objects, as well as for spatial planning with given {\it dynamic} 
geometric situation of moving objects. A full version of this paper is 
available at \cite{Pk1}.

To formulate the problem let us first present the needed notations and 
definitions, and then consider the various standard distances between 
geometric objects. We assume that the geometric objects are {\it regular} sets 
($r$-sets) in the Euclidean space $R^n$, for $n=2$ or $3$, i.e., bounded, 
closed, and semi-analytic subsets of $R^n$ \cite{Rq}. This means that, for any 
$r$-set $A$ of $R^n$, $A=\k\i (A)$, where $\i$ and $\k$ denote the 
{\it interior} and the {\it closure} of sets, respectively \cite{Rq}. The 
{\it complement} and the {\it boundary} of $A$ are denoted by $A^c$ and 
$\bd A$, respectively, and a copy of $A$ translated by a point (or a vector) 
$p$ is denoted by $A+p$. We denote by $A^\t$ a copy of $A$ rotated by an angle 
$\t$ about the origin $O$ \cite{LT}. The {\it regularized} set operations on 
two objects $A$ and $B$ in $R^n$ are defined as 
$A\otimes^*B=\k\i(A\otimes B)$, where $\otimes\in\{\cup,\cap,\bsl\}$; see 
\cite{Rq} and \cite{RV} for details. The {\it regularized} complement 
$A^{c^*}$ of $A$ is defined as $A^{c^*}=\k\i(A^c)$. The $r$-sets are not 
algebraically closed under the standard set operations, but they are closed 
under the regularized set operations \cite{Rq}. For example, the standard 
intersection $A\cap B$ of $r$-sets $A$ and $B$ needs not be regular, since its 
boundary may have {\it dangling} faces/edges and/or {\it isolated} points, but 
$A\cap^* B$ is an $r$-set.

\subsubsection*{1.2 Standard distances between geometric objects} 

Let us consider the following distances between objects $A$ and $B$ (see 
\cite{AST},\cite{H}, and \cite{LinM}):
$$d_1(B,A)=\inf_{a \in A}\inf_{b \in B}\ll a-b\ll;~~~~
d_2(B,A)=\sup_{a \in A}\sup_{b \in B}\ll a-b\ll;~~~~$$
$$h(A,B)=\sup_{a \in A}\inf_{b \in B}\ll a-b\ll;~~~~~
h(B,A)=\sup_{b \in B}\inf_{a \in A}\ll a-b\ll;~~$$
$$~~~~~~~d^*(A,B)=\inf_{a^* \in A^c}\inf_{b \in B}\ll a^*-b\ll;~~~
H(A,B)=\max\{h(A,B),h(B,A)\},$$
where $\ll \cdot \ll$ is the Euclidean norm. 
%See Figure~\ref{pr1}.
The distance $h(A,B)$ is called the {\it directed} Hausdorff distance from $A$ 
to $B$, and the distance $H(A,B)$ is called the Hausdorff distance between $A$ 
and $B$, respectively; see \cite{AST}. $H(A,B)$ is a metric on $R^n$.

These distances between $r$-sets have the following basic properties \cite{H}:
$$d_1(A,B)>0 \iff A\cap B=\es;~d_2(A,A)=\mbox{diam}(A);~H(A,B)=0 \iff A=B.$$
%$$d_2(A,A)=diameter(A);~~d^*(A,A)=0.$$
Clearly, $d^*(A,B)=d_1(A^c,B)$. Then we have $d^*(A,B)>0 \iff B\s A$.
%---------------------------------------------------------------------------
%\begin{figure}[htb]
%\begin{center}
%\input{pres_1.pstex_t}
%\caption{The standard distances between geometric objects $A$ and $B$.} 
%\label{pr1}
%\end{center}
%\end{figure}
%----------------------------------------------------------------------------

\subsubsection*{1.3 Problem formulation}

Let $A$ be an unmovable object, and let $B$ be another object, allowed only to 
translate. Let us consider the following problem:

{\bf Problem I}  For given $\l_1,\ldots,\l_6$, and the corresponding constraint
$$\nu_I(B+p,A)~=~
\Big\{\big[d_1(B+p,A)\le\l_1\big]\odot_1 
\big[d_1(B+p,A) \ge\l_2\big]\Big\}$$
$$~~~~~~~~~~~~~~~~~~~~~~~\odot_2 \Big\{\big[d_2(B+p,A)\le\l_3\big]\odot_1 
\big[d_2(B+p,A) \ge\l_4\big]\Big\}$$
$$~~~~~~~~~~~~~~~~~~~~~~~~~\odot_2\Big\{\big[d^*(B+p,A)\le\l_5\big]\odot_1 
\big[d^*(B+p,A) \ge\l_6\big]\Big\}, $$
%$$\nu_I(B+p,A)=\Big[\l_1\le d_1(B+p,A)\le\l_2\Big]\bigvee
%\Big[\l_3\le d_2(B+p,A)\le\l_4\Big]\bigvee
%\Big[\l_5\le d^*(B+p,A)\le\l_6\Big],$$
where $\odot_{1(2)}\in\{\vee,\wedge\}$, find the region $N_I(B,A)$ of all the 
feasible translations $B+p$ of $B$, in which $\nu_I(B+p,A)$ holds.
%{\bf Problem II.} For given $\e_1,\ldots,\e_6$, find the region $N_{II}(B,A)$ 
%of all placements $p$ satitfying
%\begin{eqnarray*}
%\nu_{II}(B+p,A)~=~
%\bigg\{\Big[d_1(B+p,A)\le\e_1\Big]\bigvee 
%\Big[d_1(B+p,A) \ge\e_2\Big]\bigg\} \\
%~\bigwedge~ \bigg\{\Big[d_2(B+p,A)\le\e_3\Big]\bigvee 
%\Big[d_2(B+p,A) \ge\e_4\Big]\bigg\} \\
%\bigwedge \bigg\{\Big[d^*(B+p,A)\le\e_5\Big]\bigvee 
%\Big[d^*(B+p,A) \ge\e_6\Big]\bigg\}.
%\end{eqnarray*}

Thus, our goal is to find all the feasible positions $p$ of $B$ with respect 
to $A$, under the above constraint on their relative position. This problem is 
generalization of the well known {\it Findspace} problem, formulated in 
\cite{LP}. If we let $p=O$, then the function $\nu_I(B,A)$ can be interpreted 
as the generalized Boolean {\it distance query}; see \cite{LinM} for detailes.

\subsection*{2 Preliminaries}

\subsubsection*{2.1 Minkowski operations} 
 
The {\it Minkowski sum}, and the {\it Minkowski diffference} of objects $A$ 
and $B$ are defined as
$$A\+B=\{a+b \mid a\in A, b\in B\}=\bigcup_{b\in B}(A+b),~~\mbox{and}~~
A\-B=\bigcap_{b\in B}(A+b)=(A^c\+B)^c,$$
%See Figure~\ref{pr21}(a,b).
respectively \cite{M}, \cite{Se1}. See Figure~\ref{pr21a}(a). The Minkowski 
subtraction is a {\it dual} operation of the Minkowski addition. Note that 
$A\+B=(A^c\-B)^c$.
%---------------------------------------------------------------------------
\begin{figure}[htb]
\begin{center}
\input{pres_21a.pstex_t}
\caption{(a) The Minkowski sum $\ab$, and the Minkowski difference 
$A\ominus B$ of objects $A$ and $B$. (b) The objects $\adb$, and 
$A\ominus\cB$. Here $p_1\in \bd(\adb)$, $p_2\in \bd(\adbi)$, 
$(B+p_{1,2})\dot \cap A$, and $p_3\in \bd(A\ominus \cB)$, $(B+p_3)\dot\s A$, 
respectively. (Dashed lines show an objects $B+p$, where (a) $p\in \bd A$, (b) 
$p\in \bd(\adbi)$ (resp., $p\in \bd(A\ominus \cB)$). Dotted lines show pieces 
of $\bd(\abi)$ (resp., $\bd(\adbi)$).)} 
\label{pr21a}
\end{center}
\end{figure}
%----------------------------------------------------------------------------

Let $\cB$ be the {\it reflection} of $B$ with respect to the origin $O$, i.e., 
$\cB=\{-b\mid b\in B\}$. (For notational convenience, the object $\cB$ is 
sometimes denoted by $-B$.) Then the {\it dilation}, and the {\it erosion} 
of $A$ by $B$ are defined as
$$\adb=\{a-b \mid a\in A, b\in B\}=\bigcup_{b\in B}(A-b),~~\mbox{and}~~
\aeb=\bigcap_{b\in B}(A-b)=(A^c\+\cB)^c,$$
respectively. See Figure~\ref{pr21a}(b). 

Many properties of the Minkowski operations are well known and well studied. 
See, e.g., \cite{BK}, \cite{G}, \cite{H}, \cite{LM}, \cite{M}, \cite{MMZ}, 
and \cite{Se1} for details. In this section we consider the properties of the 
Minkowski operations that we need for our purpose. 

In \cite{Ch} and \cite{G} it is shown that if an object $A$ is a convex then 
$A\- B=A\- CH(B)=A\- ext(B)$, where $CH(B)$ denote the {\it convex hull} of 
$B$, and $ext(B)$ denote the set of {\it extreme points} of $B$, i.e., the set 
of vertices of $CH(B)$. In \cite{MMZ} it is shown that the Minkowski sum 
$A\+ B$ of two $r$-sets $A$ and $B$ always results in an $r$-set, whereas the 
Minkowski difference $A\- B$ could be a non-regular set; see 
Figure~\ref{pr21a}(a). 
%This is because the Minkowski operations are defined by means of the standard 
%(but not a regularized) set operations.

Since $A\+(-(B+p))=(\adb)-p$, we have (see, e.g., \cite{BK}, \cite{GRS}, 
\cite{LP}, \cite{M}, \cite{PSG}, and \cite{SP}):
\begin{equation}
\begin{array}{lll}
(B+p)\cap A \ne \es & \iff  p\in \adb; \\
(B+p)\cap A = \es   & \iff  p\in (\adb)^c; \\
(B+p)\dot\cap A    & \iff  p\in \bd(\adbi), \\
\end{array}
\label{eq_3.1.1}
\end{equation}
where $B\dot \cap A=[(\i A\cap\i B=\es)\wedge (\bd A\cap \bd B\ne \es)]$ 
denotes the {\it outer touching} of the objects $A$ and $B$. The object 
$\adb$ is also called the {\it C-space obstacle} of $B$ relative to $A$ 
\cite{LP}. The set $\bd(\adbi)$ has been introduced in \cite{BK}, where it is 
referred to as the {\it outer envelope} of $A$ and $\cB$. In \cite{BK} it is 
shown that $\bd(\adb)\se\bd(\adbi)$, and that the set 
%that $B+p$ {\it contacts} with $A$ if and only if $p\in\bd(\adbi)$. The 
$\bd(\adbi)$ may have coincident faces/edges and/or isolated vertices, which 
are {\it removed} from the open point set $\i(\adb)$; see Figure~\ref{pr21a}.
%illustrates the case where the sets $\bd(\abi)$ and $\bd(\adbi)$ have 
%coincident edges shown by dotted lines.

From the relationships $\bd(\aeb)=\bd(A^c\+\i\cB)$ and $A\-(-(B+p))=(\aeb)-p$, 
it follows that if $A\- \cB \ne \es$, we get (see \cite{M} and \cite{Pk}):
\begin{equation}
\begin{array}{lll} 
(B+p)\s A                  & \iff p\in \aeb; \\
(B+p)\not\subset A         & \iff p\in (\aeb)^c; \\
(B+p)\dot \s A             & \iff p\in \bd(\aeb), \\
\end{array}
\label{eq_3.1.2}
\end{equation}
where $B\dot \s A=[(A^c\cap\i B=\es )\wedge (\bd A\cap \bd B\ne \es)]$ denotes 
the {\it inner touching} of $A$ and $B$. See Figure~\ref{pr21a}(b).

By observations of \cite{BK} and \cite{MMZ} we have 
$(\abi)\se \i(\ab)$; $(\ambi)\sp \i(\amb)$ and
$$\bd(\ab)\se\bd(\abi);~~\bd(\amb)=\bd(\ambi)=\bd(A^c\+\i B);$$
\begin{equation}
\ab=(\abi)\cup\bd(\abi);~~\amb= \i(\amb)\cup\bd(\amb);
\label{eq_3.1.21}
\end{equation}
\begin{equation}
(\ab)^c=(\abi)^c\bsl\bd(\abi);~~(\amb)^c=\k[(\amb)^c]\bsl\bd(\amb).
\label{eq_3.1.22}
\end{equation}
From the observations of \cite{H} it follows that, for $r$-sets, we have 
\begin{equation}
A\+\i B=\i A\+B=\abi;~A\+B=\k(\abi);~\amb=A\- \i B=\i A\-\i B.
\label{eq_3.1.23}
\end{equation}
(Clearly, for non-regular point sets, the above equalities do not necessarily 
hold.)

Let us consider the difference $\bea$. From the properties of the Minkowski 
difference (see \cite{M}) it follows that $p \in \bea$ if and only if 
$p\notin A\+ \cB^c$, and then $A \cap (B^c+p)=\es$, i.e., $A\s(B+p)$. Thus, 
$B+p$ covers $A$ if and only if $p\in \bea$. In other words,
\begin{equation}
\begin{array}{lll} 
A\s(B+p)            & \iff p\in \bea; \\
A\not\subset (B+p)  & \iff p\in (\bea)^c; \\
A\dot \s (B+p)      & \iff p\in \bd(\bea). \\
\end{array}
\label{eq_3.1.3}
\end{equation}
%See Figure~\ref{pr21b}. 
(Note that, by previous observations, $\bea$, in general, is a non-regular 
set.)
% and may have dangling faces/edges and/or isolated points.)
%---------------------------------------------------------------------------
%\begin{figure}[htb]
%\begin{center}
%\input{pres_21b.pstex_t}
%\caption{(a) The objects $A$ and $B$. (b) The object $\cB \ominus A$. Here 
%$p\in \cB \ominus A$, and $A \s (B+p)$.} 
%\label{pr21b}
%\end{center}
%\end{figure}
%----------------------------------------------------------------------------

In case where both $A$ and $B$ are allowed to translate, we have 
$(A\pm q)\+(-(B\pm p))=(\adb)\mp p \pm q$, 
$(A\pm q)\-(-(B\pm p))=(\aeb)\mp p \pm q$, and 
$(-(B\pm p))\-(A\pm q)=(\bea)\mp p \pm q$, respectively. Then all the 
relationships of (\ref{eq_3.1.1}), (\ref{eq_3.1.2}), and (\ref{eq_3.1.3}) can 
be reformulated to handle this more general form. For instance, 
$(B+p)\s (A+q)\iff (p-q) \in\aeb$. (For $\adb$ this well known fact (see, 
e.g., \cite{GRS} and \cite{Se1}) has widely been used in \cite{AB}, \cite{DM}, 
and \cite{Ml1}, for solving various containment problems.) Clearly, for $p=O$, 
we have 
$$\begin{array}{lll} 
B\cap A \ne \es & \iff O\in \adb; \\
B\s A           & \iff O\in \aeb; \\
A\s B           & \iff O\in \bea, \\
\end{array}$$
which are alternative formulations of the {\it overlapping} of the objects $A$ 
and $B$, of the {\it containment} of $B$ in $A$, and of the {\it covering} of 
$A$ by $B$, respectively.

\subsubsection*{2.2 Distances between geometric objects 
concerning their outer relative position}

The standard minimum distance $d_1(B,A)$ does not take into account the 
``amount'' of intersection between objects $A$ and $B$, since $d_1(B,A)=0$, 
for $A\cap B\ne \es$, regardless of how much they overlap. 
%Let us consider the standard minimum distance $d_1(B,A)$ between objects $A$ 
%and $B$. The distance $d_1(B,A)$ does not take into account the ``amount'' of 
%intersection between $A$ and $B$, since $d_1(B,A)=0$ for $A\cap B\ne \es$, 
%regardless of how much they overlap.
% In other words, $d_1(B,A)$, for intersecting objects, does not contain any 
%information about the degree of penetration of $A$ into $B$.
Minimum translational distance constraints that take into account penetration 
between objects have been proposed in \cite{BL}, \cite{CC}, \cite{OG}, 
\cite{P}, and \cite{St1}. The work in \cite{Pk1} consider one specific set of 
definitions of such minimum distances, since it has been defined in different 
ways. In this section we consider the translational distances, as defined in 
\cite{CC} and \cite{P}.

The {\it minimum translational} distance $MTD(A,B)$, introduced in \cite{CC}, 
is defined as
$$MTD(A,B)=\left\{\begin{array}{cc} 
-MTD^+(A,B),& ~\mbox{for}~A\cap B \ne \es; \\
~~MTD^+(A,B),& ~\mbox{otherwise}, \\
\end{array}\right.$$
where $MTD^+(A,B)=\inf \{\ll t\ll \mid A \dot \cap (B+t)\}$.

The distances $\g_{1,2}(B,A)$ between $A$ and $B$, suggested in \cite{P}, are 
defined as
$$\g_1(B,A)=\left\{\begin{array}{ccc}
\inf_{c \in \bd (\adb)}\ll c\ll ,& ~\mbox{for}~A\cap B=\es; \\
0,& ~\mbox{for}~A\dot \cap B; \\
-\inf_{c \in \bd (\adb)}\ll c\ll ,& ~\mbox{for}~A\cap B\ne\es, \\
\end{array}\right.$$
$$\g_2(B,A)=~~\sup_{c \in \bd (\adb)}\ll c\ll.~~~~~~~~~~~~
~~~~~~~~~~~~~~~~~~~~~~~~~~~~~$$
%~~~~\g_2(B,A)=~~\sup_{c \in \bd (\adb)}\ll c\ll.$$
See Figure~\ref{pr11}. The distance $\g_1(B,A)$ is defined by the above 
relationship only in case where $\bd(\adb)=\bd(\adbi)$ (see subsection 2.1). 
Then, in general, we obtain that
$$\g_1(B,A)=\left\{\begin{array}{ccc}
~~\inf_{c \in \bd (\adbi)}\ll c\ll ,& ~\mbox{for}~A\cap B=\es; \\
-\inf_{c \in \bd (\adbi)}\ll c\ll ,& ~\mbox{otherwise}. \\
\end{array}\right.$$
See Figure~\ref{pr11a}. 
%(Clearly, $\g_1(B,A)=\g_1(\i B,\i A)$.)
By the observations of subsection 2.1, we get
$\g_1(B,A)=\g_1(O,\adbi)$, and 
$$\g_1(B,A)\left\{\begin{array}{lll} 
< 0,&~ \mbox{for}~O\in \adbi; \\
=0,&~ \mbox{for}~O\in \bd (\adbi); \\
>0,&~ \mbox{for}~O\in (\adb)^c.  \\
\end{array}\right.$$
Note that in the above relationships the set $\i A$ (resp., $\i\cB$) can be 
replaced by $A$ (resp., $\cB$), since $\adbi=A\+\i\cB=\i A\+\cB$, for 
$r$-sets. See \cite{Pk1} for more details.

%---------------------------------------------------------------------------
\begin{figure}[htb]
\begin{center}
\input{pres_11.pstex_t}
\caption{The distances $\g_{1,2}(B,A)$, for various relative positions of $A$ 
and $B$, in case where $\bd(\adb)=\bd(\adbi)$. Here (a) $A\cap B\ne \es$, 
(b) $A\dot \cap B$, and (c) $A\cap B= \es$, respectively.
} 
\label{pr11}
\end{center}
\end{figure}
%----------------------------------------------------------------------------
%---------------------------------------------------------------------------
\begin{figure}[htb]
\begin{center}
\input{pres_11a.pstex_t}
\caption{The distance $\g_1(B,A)$, for various relative positions of $A$ and 
$B$, in case where $\bd(\adb)\s\bd(\adbi)$. Here (a) $A\cap B\ne\es$; $p_x$, 
$p_y$ are the values of minimal translations of $B$ in directions $x$ and $y$, 
corresponding to $\g_1(B,A)$, (b) $A\dot\cap B$, and (c) $A\cap B=\es$, 
respectively.
} 
\label{pr11a}
\end{center}
\end{figure}
%----------------------------------------------------------------------------

It can easily be shown that $\g_1(B,A)=MTD(A,B)$. Therefore we denote the 
minimum translational distance between $A$ and $B$ by $\g_1(B,A)$. 

The properties of translational distances have been well studied. (See, e.g., 
\cite{CC}, \cite{GRS}, \cite{KS}, \cite{O}, \cite{OG}, \cite{P}, \cite{Pk}, 
\cite{St1}, and \cite{SPV}. See also \cite{Pk1} for basic properties of 
$\g_{1,2}(B,A)$.) In \cite{CC}, \cite{GRS}, \cite{P}, and \cite{SPV} it is 
shown that $\g_1(B,A)$ (resp., $\g_2(B,A)$) corresponds to the minimal (resp., 
maximal) translation $B+p$ of $B$ relative to $A$ that reaches an outer 
touching $(B+p)\dot \cap A$, and that $\g_1(B,A)=d_1(B,A)$, for $A\cap B=\es$, 
and $\g_2(B,A)=\g_2(O,\adb)=d_2(B,A)$. The distances $\g_{1,2}(B,A)$ are 
invariant with respect to both rotations and translations, i.e., 
$\g_{1,2}(B^\t+p,A^\t+p)=\g_{1,2}(B,A)$; see \cite{OG}, \cite{Pk}, \cite{SPV}, 
and \cite{SY}. 

The papers \cite{P} and \cite{SPV} have considered the family of surfaces 
$\bd(\adb\+\l K)$, for $\l \ge 0$, where $\l K$ is the ball of radius $\l$ 
centered at $O$. In these works it is shown that $\g_1(B+p,A)=\l$, for 
$p\in \bd(\adb\+\l K)$. The surfaces $\bd(\adb\-|\l|K)$, with similar 
properties, for negative values of $\l$, have been defined in \cite{SPV}. 
(Note that the above relationship holds only in case where 
$\bd(\adb\+\l K)=\bd(\adbi\+\i\l K)$, for $\l> 0$, and 
$\bd(\adb)=\bd(\adbi)$, otherwise.)

\subsection*{3 Parametric families of a single object and distances between 
a point and an object}

\begin{figure}[htb]
\begin{center}
\input{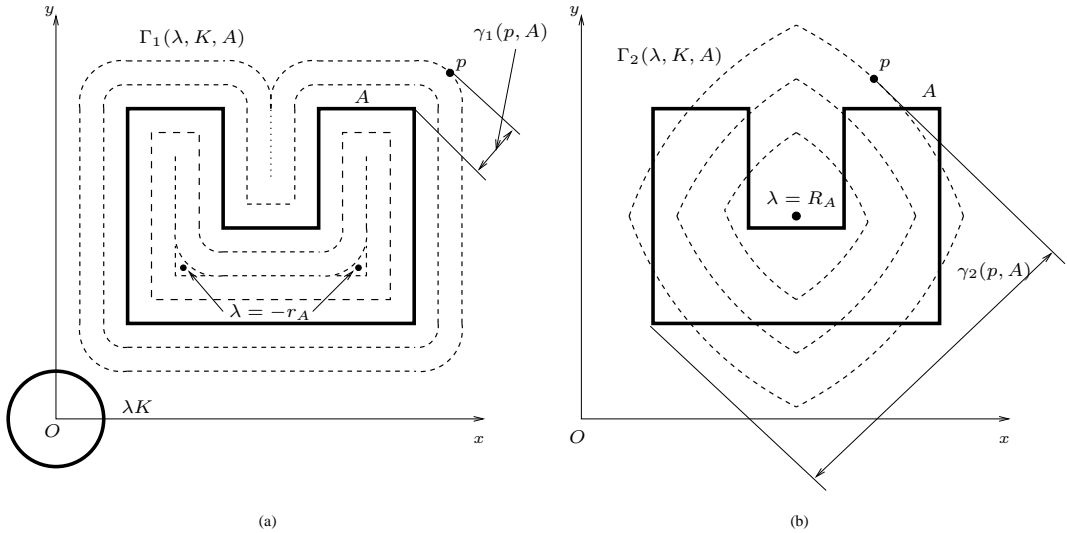}
\caption{The parametric families of objects $\G_{1,2}(\l,K,A)$, for various 
values of $\l$, and the distances $\g_{1,2}(p,A)$. Here 
$\G_2(R_A,K,A)\not\in A$, dashed curves show $\bd\G_{1,2}(\l,K,A)$, and the 
dotted line shows a piece of $\bd\G_1(\l,K,\i A)$. The objects 
$\G_1(-r_A,K,A)$ and $\G_2(R_A,K,A)$ consist of two singleton points, and a 
singleton point, respectively.}
\label{pr3b}
%\label{pr3a}
\end{center}
\end{figure}

The parametric family of objects
$$\G_1(\l,K,A)=\left\{\begin{array}{cc} 
A\+\l K,& ~\mbox{for}~\l\ge 0; \\
A\-|\l| K,& ~\mbox{for}~-r_A\le \l \le 0, 
\end{array}\right.$$
where $r_A$ is the radius of the largest inscribed ball in $A$, is called the 
{\it full parallel pencil} of the object $A$ \cite{H}, or, for $\l\ge 0$, the 
{\it offsets} of $A$ \cite{RR1}. See Figure~\ref{pr3b}(a). The object 
$\G_1(-r_A,K,A)$ is the locus of the centers of all largest inscribed balls in 
$A$, and, in general, is a (curved) face/edge. Clearly, $\G_1(-r_A,K,A)\s A$.
%Let us consider the object $S=\l K \- A$ (see \cite{H} and \cite{Pk2}), and 
%let $R_A$ be the radius of the smallest circumscribed ball of $A$. Then from 
%the definition of Minkowski difference it follows that $S=\es$ if and only if 
%$\l<R_A$.

A second parametric family of objects
$$\G_2(\l,K,A)=\l K \- A,~\mbox{for}~\l \ge R_A,~~~~~~~~~~~~~~~$$
%$\G_2(\l,K,A)=\l K\- A$, for $\l\ge R_A$, 
where $R_A$ is the radius of the smallest circumscribed ball of $A$, has been 
introduced in \cite{Pk2} and \cite{Pk3}. See Figure~\ref{pr3b}(b). From the 
definition of Minkowski operations, it follows that $\G_1(\l,K,A)=\es$ if and 
only if $\l< -r_A$, and $\G_2(\l,K,A)=\es$ if and only if $\l<R_A$. Since 
$\l K$ is convex, it follows that $\G_2(\l,K,A)$ is convex, for any bounded 
$A$, and $\G_2(\l,K,A)=\G_2[\l,K,CH(A)]=\G_2[\l,K,ext(A)]$; see 
subsection 2.1. The object $\G_2(R_A,K,A)$ is a singleton point, which is the 
center of the (unique) smallest enclosing ball of $A$. In general, the point 
of $\G_2(R_A,K,A)$ needs not be in $A$; see Figure~\ref{pr3b}(b). From the 
definitions of $\G_1(\l,K,A)$ and $\G_2(\l,K,A)$ it follows that  
$\G_2(\l,K,A)\se\G_1(\l,K,A)$, for $\l\ge R_A$. (An equality holds in case 
where $A$ is a singleton point.)
%---------------------------------------------------------------------------
%----------------------------------------------------------------------------

Let us consider the translational distances $\g_{1,2}(p,A)$ between a point 
$p$ and an object $A$:
$$\g_1(p,A)=\left\{\begin{array}{ccc}
\inf_{a \in \bd A}\ll p-a\ll ,& ~\mbox{for}~p\in A^c; \\
0,& ~\mbox{for}~p\in \bd A; \\
-\inf_{a \in \bd A}\ll p-a\ll ,& ~\mbox{for}~p\in \i A, \\
\end{array}\right.$$
$$\g_2(p,A)=~~~~~\sup_{a \in \bd A}\ll p-a\ll .~~~~~~~~~~~~
~~~~~~~~~~~~~~~~~~~~$$
%~~~~~\g_2(p,A)=\sup_{a \in \bd A}\ll p-a\ll.$$
%Note that $\g_1(p,A)=d_1(p,A)$, for $p\in A^c$, $\g_2(p,A)=d_2(p,A)$, and 
%$\g_1(p,A)=-\g_1(p,A^c)$.
To lack of space, all proofs here and in the following sections are omitted. 
The proofs of the observations, lemmas, and theorems are presented in 
\cite{Pk1}.

The distances $\g_{1,2}(p,A)$ have the following properties:
\begin{observation} \label{ob_5.1}
(a) For $\l\ge 0$ we have
$$\g_1(p,A)\left\{\begin{array}{lll}
<\l,~\mbox{for}~p \in \G_1(\l,K,\i A);\\
=\l,~\mbox{for}~p \in \bd\G_1(\l,K,\i A);\\
>\l,~\mbox{for}~p \in [\G_1(\l,K,A)]^c,\\
\end{array}\right.
%\mbox{and}~ 
\g_1(p,A)\left\{\begin{array}{lll}
\le\l, ~\mbox{for}~p \in \G_1(\l,K,A);\\
\ne\l, ~\mbox{for}~p \in [\bd\G_1(\l,K,\i A)]^c;\\
\ge\l, ~\mbox{for}~p \in [\G_1(\l,K,\i A)]^c.\\
\end{array}\right.$$
(b) For $-r_A\le\l\le 0$ we have
$$\g_1(p,A)\left\{\begin{array}{lll}
<\l, ~\mbox{for}~p \in \i\G_1(\l,K,A);\\
=\l, ~\mbox{for}~p \in \bd\G_1(\l,K,A);\\
>\l, ~\mbox{for}~p \in [\G_1(\l,K,A)]^c,\\
\end{array}\right.
%\mbox{and}~ 
\g_1(p,A)\left\{\begin{array}{lll}
\le\l, ~\mbox{for}~p \in \G_1(\l,K,A);\\
\ne\l, ~\mbox{for}~p \in [\bd\G_1(\l,K,A)]^c;\\
\ge\l, ~\mbox{for}~p \in \k[\G_1(\l,K,A)]^c.\\
\end{array}\right.$$
\end{observation}
\begin{observation} \label{ob_5.2}
For $\l\ge R_A$ we have
$$\g_2(p,A)\left\{\begin{array}{lll}
<\l, ~\mbox{for}~p \in \i \G_2(\l,K,A);\\
=\l, ~\mbox{for}~p \in \bd\G_2(\l,K,A);\\
>\l, ~\mbox{for}~p \in [\G_2(\l,K,A)]^c,\\
\end{array}\right.
%\mbox{and}~ 
\g_2(p,A)\left\{\begin{array}{lll}
\le\l, ~\mbox{for}~p \in \G_2(\l,K,A);\\
\ne\l, ~\mbox{for}~p \in [\bd \G_2(\l,K,A)]^c;\\
\ge\l, ~\mbox{for}~p \in \k [{\G_2}(\l,K,A)]^c.\\
\end{array}\right.$$
\end{observation}
It is clear that $\inf_{p\in R^n}\{\g_{1(2)}(p,A)\}=-r_A(R_A)$. Generally, 
for $i=1,2$, we get
\begin{equation}
\g_i(p,A)\left\{\begin{array}{lll}
\le\l, ~\mbox{for}~p \in \G_i(\l,K,A);\\
>\l, ~\mbox{for}~p \in [{\G_i}(\l,K,A)]^c.\\
\end{array}\right.
\label{eq_3.4}
\end{equation}
The topological properties of families $\G_{1,2}(\l,K,A)$ have been studied in 
\cite{Pk1}.

\subsection*{4 Correspondence between distances and the parametric families}

Let us consider the translational distance $\o(B,A)$ between the objects $A$ 
and $B$. We say that the distance $\o(B,A)$ {\it corresponds} to the 
parametric family of objects $\O(\l,K,B,A)$ if and only if 
$\o(B+p,A)\le\l$, for $p \in \O(\l,K,B,A)$. 
%$$\o(B+p,A)\left\{\begin{array}{lll} 
%\le\l, ~\mbox{for}~p \in \O(\l,K,B,A);\\
%>\l, ~\mbox{for}~p \in [\O(\l,K,B,A)]^c.\\
%\end{array}\right.$$
(Clearly, $\o(B+p,A)>\l$, for $p \in [\O(\l,K,B,A)]^c$, and  $\o(B,A)\le\l$, 
for $O\in\O(\l,K,B,A)$.) The correspondence between $\o(B,A)$ and 
$\O(\l,K,B,A)$ is denoted by $\o(B,A)\sim\O(\l,K,B,A)$.
%
%\subsubsection*{4.1 The special case}
%
%Let us consider the special case where an object $B$ is the origin $O$. Then 

In special case where an object $B$ is the origin $O$, the distance function 
$\o(B+p,A)$ is reduced to the distance $\o(O+p,A)=\o(p,A)$ between a point $p$ 
and an object $A$, and the family of objects $\O(\l,K,B,A)$ is reduced to 
$\O(\l,K,O,A)=\O(\l,K,A)$.

Consider the families of objects $\G_{1,2}(\l,K,A)$, and the distances 
$\g_{1,2}(O+p,A)$.
\begin{lemma} \label{l1}
For $i=1,2$ we have: (a) $\g_i(O,A)\sim \G_i(\l,K,A)$. 
(b) $\g_i(\pm p,A)=\g_i(O,A\mp p)$. 
\end{lemma} 
Follow from the relationship (\ref{eq_3.4}), and since 
$A\+\{\pm\check p\}=A\mp p$.

Since $\mbox{diam}(A)=2R_A$, it can easily be shown that
$\g_2(p,A)=\mbox{diam}(\{p\}\cup A)>\l$ if and only if 
$p\in [\G_2(\l,K,A)]^c$, for $\l\ge 2R_A$. Then we get 
$\mbox{diam}(\{O\}\cup A)\sim \G_2(\l,K,A)$, for $\l\ge 2R_A$. (Note that 
$\g_2(p,A)\le\mbox{diam}(\{p\}\cup A)=\mbox{diam}(A)=\l$  if and only if 
$p\in\G_2(\l,K,A)$, for $\l\le 2R_A$.)

The properties of the distances $\g_{1,2}(p,A)$ and their corresponding 
families $\G_{1,2}(\l,K,A)$ in case where $A=\bigcup_{j=1}^n A_j$ and/or 
$A=\bigcap_{j=1}^n A_j$, for $A\ne\es$, have been studied in \cite{Pk1}.
%
%\subsubsection*{4.2 The general case}
%

In general case where $B$ is a geometric object, but not a single point, we 
have
\begin{lemma} \label{l2}
(a) $\g_i(B+p,A)=\g_i(B,A-p)=\g_i(p,\adbi)$, for $i=1,2$. 
(b) $\g_i(p,\adbi)=\g_i[B+\a \cdot p,A-(1-\a)\cdot p]$, for $0\le\a\le 1$. 
(c) $\g_i(B^\t \pm p,A^\t \pm q)=\g_i(B^\t \mp q,A^\t \mp p)=
\g_i[\pm p\mp q,(\adbi)^\t]$, for $i=1,2$.
\end{lemma} 

{\bf Remark 1} For $i=2$, the sets $\i A$ and/or $\i B$ can be replaced by 
$A$ and/or $B$, respectively, since 
$\sup\{\ll c\ll \mid c\in \bd(\adbi)\}=\sup\{\ll c\ll \mid c\in \bd(\adb)\}$, 
i.e., $\g_2(p,\adbi)=\g_2(p,\adb)$.

{\bf Remark 2} By the first of relationships ({\ref{eq_3.1.23}}), we have 
$\g_i(B,A)=\g_i(B^*,A^*)$, for $i=1,2$, where $A^*\in\{A,\i A\}$ and 
$B^*\in\{B,\i B\}$, respectively.

Let us first consider the parametric family of objects
$$\G_1(\l,K,B,A)=\left\{\begin{array}{cc}
(\adb)\+\l K,& ~\mbox{for}~\l\ge 0; \\
(\adb)\- |\l|K,& ~\mbox{for}~-r_{\adb}\le \l \le 0,
\end{array}\right.$$
(see Figure~\ref{pr78a}(a)). Since $\G_1(\l,K,B,A)=\G_1(\l,K,\adb)$, the 
object $\G_1(-r_{\adb},K,B,A)$ is the locus of the centers of all largest 
inscribed balls in $\adb$.
\begin{observation} \label{ob_6.2.1}
(a) For $\l\ge 0$ we have
$$\g_1(B+p,A)\left\{\begin{array}{lll}
\le\l,~\mbox{for}~p \in \G_1(\l,K,B,A);\\
<\l,~\mbox{for}~p \in \G_1(\l,K,B,\i A);\\
=\l,~\mbox{for}~p \in \bd\G_1(\l,K,B,\i A);\\
\ge\l,~\mbox{for}~p \in [\G_1(\l,K,B,\i A)]^c.\\
%>\l,~\mbox{for}~p \in [\G_1(\l,K,B,A)]^c.\\
\end{array}\right.$$
%\mbox{and}~
%\g_1(B+p,A)\left\{\begin{array}{lll}
%\le\l,~\mbox{for}~p \in \G_1(\l,K,B,A);\\
%\ne\l,~\mbox{for}~p \in [\bd\G_1(\l,K,B,\i A)]^c;\\
%\ge\l,~\mbox{for}~p \in [\G_1(\l,K,B,\i A)]^c.\\
%\end{array}\right.$$
(b) For $-r_{\adbi}\le\l<0$ we have
$$~~~~\g_1(B+p,A)\left\{\begin{array}{lll}
\le\l, ~\mbox{for}~p \in \G_1(\l,\i K,\i B,\i A);\\
<\l, ~\mbox{for}~p \in \i\G_1(\l,\i K,\i B,\i A);\\
=\l, ~\mbox{for}~p \in \bd\G_1(\l,\i K,\i B,\i A);\\
\ge\l, ~\mbox{for}~p \in \k[\G_1(\l,\i K,\i B,\i A)]^c.\\
%>\l, ~\mbox{for}~p \in [\G_1(\l,\i K,\i B,\i A)]^c.\\
\end{array}\right.$$
%\mbox{and}~
%\g_1(B+p,A)\left\{\begin{array}{lll}
%\le\l, ~\mbox{for}~p \in \G_1(\l,\i K,\i B,\i A);\\
%\ne\l, ~\mbox{for}~p \in [\bd\G_1(\l,\i K,\i B,\i A)]^c;\\
%\ge\l, ~\mbox{for}~p \in \k[\G_1(\l,\i K,\i B,\i A)]^c.\\
%\end{array}\right.$$
\end{observation}
See Figures  \ref{pr78a}(a) and \ref{pr78b}(a). Hence, the set
$$P(\l,K,B,A)=\big\{p\mid \g_1(B+p,A)=\l\big\}=
\left\{\begin{array}{lll}
\bd\G_1(\l,K,B,\i A)~\mbox{for}~\l\ge 0 ;\\
\bd\G_1(\l,\i K,\i B,\i A),~\mbox{for}~-r_{\adbi}\le\l<0,\\
\end{array}\right.$$
is the surface of the function $\g_1(B+p,A)$, and 
$\inf_{p\in R^n}\{\g_1(B+p,A)\}=-r_{\adbi}$. 
%Since, for $\l\ge 0$, holds $\G_1(\l,K,B,A)=\k\G_1(\l,K,B,\i A)$, then 
%from the Observation~\ref{ob_6.2.1} it follows that
%$$\g_1(B+p,A)~~\left\{\begin{array}{lll}
%\le\l,~\mbox{for}~p \in \G_1(\l,K,B,A);\\
%\ne\l,~\mbox{for}~p \in [\bd\G_1(\l,K,B,\i A)]^c;\\
%\ge\l,~\mbox{for}~p \in [\G_1(\l,K,B,\i A)]^c,\\
%\end{array}\right.$$
%for $\l\ge 0$, and 
%$$\g_1(B+p,A)~~\left\{\begin{array}{lll}
%\le\l, ~\mbox{for}~p \in \G_1(\l,\i K,\i B,\i A);\\
%\ne\l, ~\mbox{for}~p \in [\bd\G_1(\l,\i K,\i B,\i A)]^c;\\
%\ge\l, ~\mbox{for}~p \in \k[\G_1(\l,\i K,\i B,\i A)]^c,\\
%\end{array}\right.$$
%for $-r_{\adbi}\le\l<0$, respectively. See Figures \ref{pr78a}(a) and 
%\ref{pr78b}(a). 
Thus, we get
\begin{theorem} \label{t_6.2.1}
$\g_1(B,A)\sim\left\{\begin{array}{lll}
\G_1(\l,K,B,A),~\mbox{for}~\l\ge0;\\
\G_1(\l,\i K,\i B,\i A),~\mbox{for}~-r_{\adbi}\le\l<0.\\
\end{array}\right.$
\end{theorem}
Note that in case where $\l\ge 0$ the Theorem \ref{t_6.2.1} has also been 
proved in \cite{P} and \cite{SPV}. 

Denote by $d_P(A,B)$ the penetration depth of $A$ and $B$ (see \cite{OG}). In 
\cite{Pk1} it is shown that $\g_1(B,A)=-d_P(\i A,\i B)$, for $A\cap B\ne\es$.
Then we have $d_P(A,B)\sim \G_1(\l,K,B,A)$, for $-r_{\adb}\le\l<0$.

Let us next consider the parametric family of objects
$$\G_2(\l,K,B,A)=\l K\- (\adb),~\mbox{for}~\l \ge R_{\adb},~~~~~~~~~~~~~~~~$$
proposed in \cite{Pk} and \cite{Pk2}; see Figures~\ref{pr78a}(b) and 
\ref{pr78b}(b). Since $\G_2(\l,K,B,A)=\G_2(\l,K,\adb)$, the object 
$\G_2(R_{\adb},K,B,A)$ is a singleton point, which is the center of the 
(unique) smallest enclosing ball of $\adb$. The object $\G_2(\l,K,B,A)$ is 
convex, for any bounded $A$ and $B$, and 
$\G_2(\l,K,B,A)=\G_2[\l,K,CH(\adb)]=\G_2[\l,K,ext(\adb)]$. 
%---------------------------------------------------------------------------
\begin{figure}[htb]
\begin{center}
\input{pres_78a.pstex_t}
\caption{The objects $\G_{1,2}(\l_{1,2},K,B,A)$. Here 
$\g_{1,2}(B+p,A)=\l_{1,2}$ and $\l_1>0$. A dotted line shows a piece of 
$\bd\G_1(\l_1,K,B,\i A)$, and $\g_1(B+q,A)=\l_1$, for 
$q\in\bd\G_1(\l_1,K,B,\i A)$.} 
\label{pr78a}
\end{center}
\end{figure}
%----------------------------------------------------------------------------
%---------------------------------------------------------------------------
\begin{figure}[htb]
\begin{center}
\input{pres_78b.pstex_t}
\caption{The objects (a) $\G_1(\l_1,\i K,\i B,\i A)$ and (b) 
$\G_2(\l_2,K,B,A)$. Here $\g_{1,2}(B+p,A)=\l_{1,2}$ and $\l_1<0$. A dashed 
lines show $\bd(\adbi)$.
} 
\label{pr78b}
\end{center}
\end{figure}
%----------------------------------------------------------------------------
\begin{observation} \label{ob_6.2.2}
For $\l\ge R_{\adb}$ we have 
$$\g_2(B+p,A)\left\{\begin{array}{lll}
\le\l, ~\mbox{for}~p \in \G_2(\l,K,B,A);\\
<\l, ~\mbox{for}~p \in \i\G_2(\l,K,B,A);\\
=\l, ~\mbox{for}~p \in \bd\G_2(\l,K,B,A);\\
\ge\l, ~\mbox{for}~p \in \k [\G_2(\l,K,B,A)]^c.\\
%>\l, ~\mbox{for}~p \in [\G_2(\l,K,B,A)]^c.\\
\end{array}\right.$$
%\mbox{and}~
%\g_2(B+p,A)\left\{\begin{array}{lll}
%\le\l, ~\mbox{for}~p \in \G_2(\l,K,B,A);\\
%\ne\l, ~\mbox{for}~p \in [\bd\G_2(\l,K,B,A)]^c;\\
%\ge\l, ~\mbox{for}~p \in \k [\G_2(\l,K,B,A)]^c.\\
%\end{array}\right.$$
\end{observation}
(Clearly, $\inf_{p\in R^n}\{\g_2(B+p,A)\}=R_{\adb}$.) Then we get the 
following:
\begin{theorem} \label{t1}
$\g_2(B,A)\sim \G_2(\l,K,B,A)$, for $\l\ge R_{\adb}$.
\end{theorem}

By previous observations, we also obtain that 
$\g_2(B+p,A)=\mbox{diam}((B+p)\cup A)>\l$ if and only if 
$p\in [\G_2(\l,K,B,A)]^c$, for $\l\ge 2R_{\adb}$. Hence, 
$\mbox{diam}(B\cup A)\sim \G_2(\l,K,B,A)$, for $\l\ge 2R_{\adb}$.

%The objects of $\G_i(\l,K,B,A)$, for $i=1,2$, have the following helpful 
%properties \cite{Pk}:
%$$\G_i(\l,K,B,A)=-\G_i(\l,K,A,B);$$
%$$\G_i[\l,K,B+\a\cdot p,A-(1-\a)\cdot p]
%=\G_i(\l,K,B,A)-p,~\mbox{for}~0\le \a\le 1;$$
%$$\G_i(\l,K,B^{\t}\pm p,A^{\t}\pm q)
%=\G_i(\l,K,B^{\t}\mp q,A^{\t}\mp p)=[\G_i(\l,K,B,A)]^{\t}\mp p \pm q.$$
The additional (topological and set theoretic)  properties of the distances 
$\g_{1,2}(B,A)$ and the families $\G_{1,2}(\l,K,B,A)$ have been studied in 
\cite{Pk1}.

\subsection*{5 Distances between geometric objects concerning their inner 
relative position} 

The duality of the Minkowski operations provide estimating the {\it inner} 
relative position of geometric objects. 

\subsubsection*{5.1 Distances concerning the containment of objects} 

\begin{figure}[htb]
\begin{center}
\input{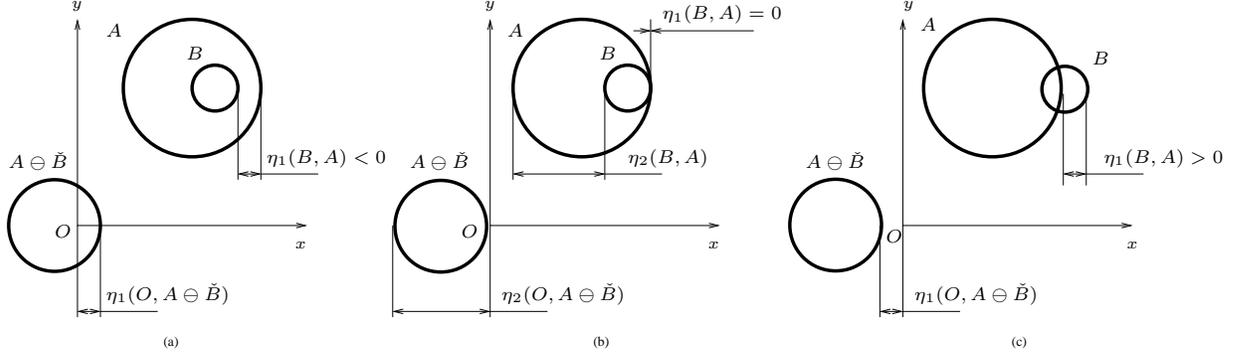}
\caption{The distances $\eta_{1,2}(B,A)$, for various relative positions of 
$A$ and $B$. Here (a) $ B\s A$, (b) $B\dot\s A$, and (c) $B \not\subset A$, 
respectively.} 
\label{pr12}
\end{center}
\end{figure}

Let us consider the following translational distances, introduced in 
\cite{Pk} and \cite{Pk2}:
$$\eta_1(B,A)=\left\{\begin{array}{cc} 
\inf_{c \in \bd (\aeb)}\ll c\ll ,& ~\mbox{for}~ B\not\subset A; \\
0,& ~\mbox{for}~B \dot \s A; \\
-\inf_{c \in \bd (\aeb)}\ll c\ll ,& ~\mbox{for}~B\s A, \\
\end{array}\right.$$
$$\eta_2(B,A)=~~~\sup_{c \in \bd (\aeb)}\ll c\ll .~~~~~~~~~
~~~~~~~~~~~~~~~~~~~~~~~~~$$
See Figure~\ref{pr12}. (The distances $\eta_{1,2}(B,A)$ are defined only in 
case where $\aeb\ne\es$.)
%---------------------------------------------------------------------------
%----------------------------------------------------------------------------

The properties of the distances $\eta_{1,2}(B,A)$ have been studied in 
\cite{Pk} and \cite{Pk3}. The distance $\eta_1(B,A)$ (resp., $\eta_2(B,A)$) 
corresponds to the minimal (resp., maximal) translation $B+p$ of $B$ relative 
to $A$ that reaches an inner touching $(B+p)\dot \s A$. Since 
$\bd(\aeb)=\bd(A^c\+\i\cB)$, we have $\eta_1(B,A)=-\g_1(B,A^c)$ and 
$\eta_1(B,A)=-d^*(A,B)$, for $B\s A$.
%The distances $\eta_i(B,A)$, 
%for $i=1,2$, have also the following simple properties:
%$$\eta_1(B,A)=-d^*(A,B),{eq_3.1.21}~\mbox{for}~B\s A;$$
%$$\eta_i(B,A)=\eta_i(O,\aeb);~~
%\eta_i(B^\t \pm p,A^\t \pm p)=\eta_i(B,A);~~
%~\mbox{for}~i=1,2.$$
%\eta_i(B,A)=\eta_i(\i B,A)=\eta_i(\i B,\i A).$$
%By the third of relationships (\ref{eq_3.1.23}), we get 
%$\eta_i(B,A)=\eta_i(\i B,A)=\eta_i(\i B,\i A)$, for $i=1,2$.
\begin{lemma} \label{l3}
(a) $\eta_i(p,A)=\g_i(p,A)$, for $i=1,2$. 
(b) $\eta_i(B+p,A)=\eta_i(B,A-p)=\eta_i(p,\aeb)$. 
(c) $\eta_i[B+\a \cdot p,A-(1-\a)\cdot p]=\eta_i(p,\aeb)$, for $0\le\a\le 1$.
(d) $\eta_i(B^\t \pm p,A^\t \pm q)=\eta_i(B^\t \mp q,A^\t \mp p)=
\eta_i[\pm p\mp q,(\aeb)^\t]$, for $i=1,2$.
\end{lemma} 
It can easily be shown that
$$\eta_1(B,A)\left\{\begin{array}{lll} 
<0,&~ \mbox{for}~O\in \i (\aeb); \\
=0,&~ \mbox{for}~O\in \bd (\aeb); \\
>0,&~ \mbox{for}~O\in (\aeb)^c.  \\
\end{array}\right.$$
See Figure~\ref{pr12}. The distances $\eta_{1,2}(B,A)$ have used in \cite{Pk2} 
and \cite{Pk3} to describe the constraints on the relative position of 
objects in containment problems.

\begin{figure}[htb]
\begin{center}
\input{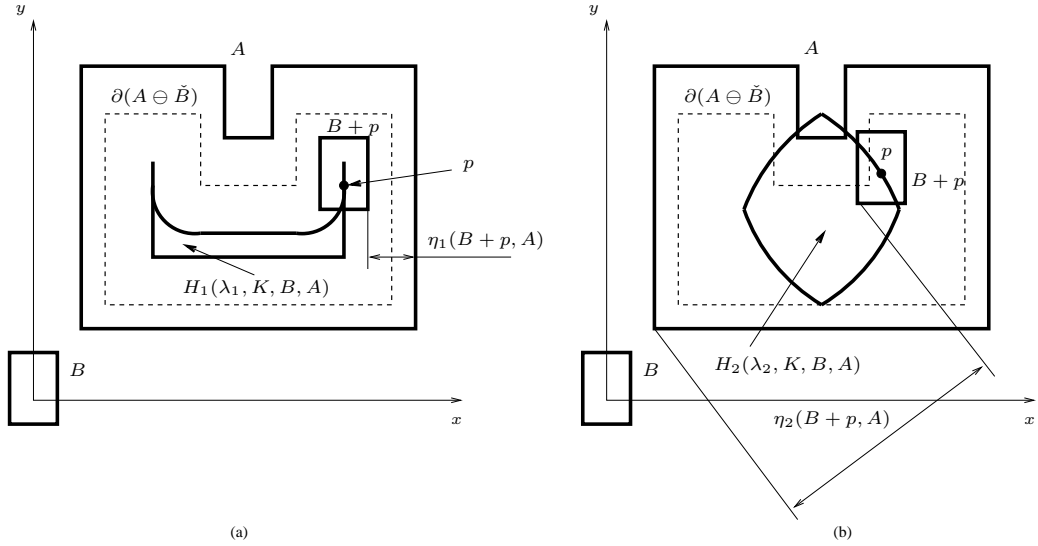}
\caption{The objects $H_{1,2}(\l_{1,2},K,B,A)$. Here 
$\eta_{1,2}(B+p,A)=\l_{1,2}$ and $\l_1<0$.} 
\label{pr910b}
\end{center}
\end{figure}

Let us consider the following parametric families of objects \cite{Pk}, 
\cite{Pk2}:
$$H_1(\l,K,B,A)=\left\{\begin{array}{cc} 
(\aeb)\+\l K,& ~\mbox{for}~\l\ge 0; \\
(\aeb)\- |\l| K,& ~\mbox{for}~-r_{\aeb}\le \l \le 0, 
\end{array}\right.$$
$$H_2(\l,K,B,A)=~~~~~~~\l K\- (\aeb),~~~~~~~~~\mbox{for}~\l \ge R_{\aeb},~~
~~~~~~~$$
See Figure~\ref{pr910b}.
%---------------------------------------------------------------------------
%----------------------------------------------------------------------------
%\input{pres_9.pstex_t} \caption{An object $H_1(\l,B,A)$, for $\l=\e_1$.} 
%\input{pres_10.pstex_t} \caption{An object $H_2(\l,B,A)$, for $\l=\e_2$.} 
\begin{observation} \label{ob_7.1.1}
(a) For $\l>0$ we have
$$\eta_1(B+p,A)\left\{\begin{array}{lll}
\le\l,~\mbox{for}~p \in H_1(\l,K,B,A);\\
<\l,~\mbox{for}~p \in H_1(\l,\i K,B,A);\\
=\l,~\mbox{for}~p \in \bd H_1(\l,\i K,B,A);\\
\ge\l,~\mbox{for}~p \in [H_1(\l,\i K,B,A)]^c.\\
%>\l,~\mbox{for}~p \in [H_1(\l,K,B,A)]^c.\\
\end{array}\right.$$
%\eta_1(B+p,A)\left\{\begin{array}{lll}
%\le\l,~\mbox{for}~p \in H_1(\l,K,B,A);\\
%\ne\l,~\mbox{for}~p \in [\bd H_1(\l,\i K,B,A)]^c;\\
%\ge\l,~\mbox{for}~p \in [H_1(\l,\i K,B,A)]^c,\\
%\end{array}\right.$$
(b) For $i=1,2$, $-r_{\aeb}\le\l_1\le 0$, and $\l_2\ge R_{\aeb}$, 
respectively, we have
$$~~~~\eta_i(B+p,A)\left\{\begin{array}{lll}
\le\l_i, ~\mbox{for}~p \in H_i(\l_i,K,B,A);\\
<\l_i, ~\mbox{for}~p \in \i H_i(\l_i,K,B,A);\\
=\l_i, ~\mbox{for}~p \in \bd H_i(\l_i,K,B,A);\\
\ge\l_i, ~\mbox{for}~p \in \k[H_i(\l_i,K,B,A)]^c.\\
%>\l_i, ~\mbox{for}~p \in [H_i(\l_i,K,B,A)]^c.\\
\end{array}\right.$$
%\eta_i(B+p,A)\left\{\begin{array}{lll}
%\le\l_i, ~\mbox{for}~p \in H_i(\l_i,K,B,A);\\
%\ne\l_i, ~\mbox{for}~p \in [\bd H_i(\l_i,K,B,A)]^c;\\
%\ge\l_i, ~\mbox{for}~p \in \k[H_i(\l_i,K,B,A)]^c,\\
%\end{array}\right.$$
\end{observation}
%Then from the Observation~\ref{ob_7.1.1} it follows that
%$$\eta_1(B+p,A)~~\left\{\begin{array}{lll}
%\le\l,~\mbox{for}~p \in H_1(\l,K,B,A);\\
%\ne\l,~\mbox{for}~p \in [\bd H_1(\l,\i K,B,A)]^c;\\
%\ge\l,~\mbox{for}~p \in [H_1(\l,\i K,B,A)]^c,\\
%\end{array}\right.$$
%for $\l>0$, and 
%$$\eta_i(B+p,A)~~\left\{\begin{array}{lll}
%\le\l_i, ~\mbox{for}~p \in H_i(\l_i,K,B,A);\\
%\ne\l_i, ~\mbox{for}~p \in [\bd H_i(\l_i,K,B,A)]^c;\\
%\ge\l_i, ~\mbox{for}~p \in \k[H_i(\l_i,K,B,A)]^c,\\
%\end{array}\right.$$
%for $i=1,2$, $-r_{\aeb}\le\l_1\le 0$, and $\l_2\ge R_{\aeb}$, respectively.
%------------------------------------------------
%%%%\begin{observation} \label{ob_7.1.2}
%%%%For $\l\ge R_{\aeb}$ holds
%%%%$$\eta_2(B+p,A)~~\left\{\begin{array}{lll}
%%%%<\l,~\mbox{for}~p \in \i H_2(\l,K,B,A);\\
%%%%=\l,~\mbox{for}~p \in \bd H_2(\l,K,B,A);\\
%%%%>\l,~\mbox{for}~p \in [H_2(\l,K,B,A)]^c.\\
%%%%\end{array}\right.$$
%%%%\end{observation}
%%%%Follows from the Observation~\ref{ob_6.2.2}, and since 
%%%%$H_2(\l,K,B,A)=\G_2(\l,K,\aeb)$. We also get  
%%%%$$\eta_2(B+p,A)~~\left\{\begin{array}{lll}
%%%%\le\l,~\mbox{for}~p \in H_2(\l,K,B,A);\\
%%%%\ne\l,~\mbox{for}~p \in [\bd H_2(\l,K,B,A)]^c;\\
%%%%\ge\l,~\mbox{for}~p \in \k [H_2(\l,K,B,A)]^c.\\
%%%%\end{array}\right.$$
%------------------------------------------------
Then $\inf_{p\in R^n}\{\eta_{1(2)}(B+p,A)\}=-r_{\aeb}(R_{\aeb})$. Thus, by 
above, we obtain the following:
\begin{theorem} \label{t2}
$\eta_i(B,A)\sim H_i(\l,K,B,A)$, for $i=1,2$.
\end{theorem}
The additional properties of the distances $\eta_{1,2}(B,A)$ and the families 
$H_{1,2}(\l,K,B,A)$ have been studied in \cite{Pk1}.

\subsubsection*{5.2 Distances concerning the covering of objects} 

In this section we introduce the distances $\dl_{1,2}(B,A)$ involving the 
{\it covering} of $A$ by $B$:
$$\dl_1(B,A)=\left\{\begin{array}{cc} 
\inf_{c \in \bd (\bea)}\ll c\ll ,& ~\mbox{for}~ A\not\subset B; \\
0,& ~\mbox{for}~A \dot \s B; \\
-\inf_{c \in \bd (\bea)}\ll c\ll ,& ~\mbox{for}~A\s B, \\
\end{array}\right.$$
$$\dl_2(B,A)=~~~\sup_{c \in \bd (\bea)}\ll c\ll .~~~~~~~~~~~~~~~~~
~~~~~~~~~~~~~~~~~~$$
See Figure~\ref{pr12a}. (The distances $\dl_{1,2}(B,A)$ are defined only in 
case where $\bea\ne\es$.)
%---------------------------------------------------------------------------
\begin{figure}[htb]
\begin{center}
\input{pres_12a.pstex_t}
\caption{The distances $\dl_{1,2}(B,A)$, for various relative positions of $A$ 
and $B$. Here (a) $A\s B$, (b) $A\dot\s B$, and (c) $A\not\subset B$, 
respectively.} 
\label{pr12a}
\end{center}
\end{figure}
The properties of the distances $\dl_{1,2}(B,A)$ and $\eta_{1,2}(B,A)$ are 
similar. The distance $\dl_1(B,A)$ (resp., $\dl_2(B,A)$) corresponds to the 
minimal (resp., maximal) translation $B+p$ of $B$ relative to $A$ that reaches 
an inner touching $A\dot\s (B+p)$. The following helpful properties of 
$\dl_{1,2}(B,A)$ hold: 
%$$\dl_1(B,A)=-\g_1(B^c,A);~\dl_1(B,A)=-d^*(B,A),~\mbox{for}~A\s B;$$
%$$\dl_i(B,A)=\eta_i(\cA,\cB)=-\g_1(B^c,A);~~
%\dl_i(B,A)=\dl_i(B,\i A)=\dl_i(\i B,\i A);$$
%\dl_i(B,A)=\dl_i(B\- \cA,O)=\g_i(O,\bea);~~
%\dl_i(B^\t \pm p,A^\t \pm p)=\dl_i(B,A),~\mbox{for}~i=1,2.$$
%$$\dl_i(B,A)=\eta_i(\cA,\cB)=-\g_1(B^c,A);~
%\dl_i(B,A)=\dl_i(B\- \cA,O)=\g_i(O,\bea);~\mbox{for}~i=1,2.$$
$\dl_1(B,A)=-\g_1(B^c,A)$ and $\dl_i(B,A)=\eta_i(\cA,\cB)$, for $i=1,2$.
\begin{lemma} \label{l5}
(a) $\dl_i(B+p,A)=\dl_i(B,A-p)=\dl_i(B\- \cA,-p)=\g_i(p,\bea)$, for $i=1,2$. 
(b) $\dl_i[B+\a\cdot p,A-(1-\a)\cdot p]=\dl_i(B\-\cA,-p)$, for $0\le \a \le 1$.
(c) $\dl_i(B^\t \pm p,A^\t \pm q)=\dl_i(B^\t \mp q,A^\t \mp p)=
\dl_i[(B\-\cA)^\t,\mp p\pm q]=\g_i[\pm p\mp q,(\bea)^\t]$, for $i=1,2$.
\end{lemma} 
It can easily be shown that
$$\dl_1(B,A)\left\{\begin{array}{lll} 
<0,&~ \mbox{for}~O\in \i (\bea); \\
=0,&~ \mbox{for}~O\in \bd (\bea); \\
>0,&~ \mbox{for}~O\in (\bea)^c.  \\
\end{array}\right.$$
See Figure~\ref{pr12a}. The distances $\dl_{1,2}(B,A)$ can be used to 
formalize the constraints on the relative position of objects in covering 
problems.

Let us consider the parametric families of objects
$$\Dl_1(\l,K,B,A)=\left\{\begin{array}{cc} 
(\bea)\+\l K,& ~\mbox{for}~\l\ge 0; \\
(\bea)\- |\l| K,& ~\mbox{for}~-r_{\bea}\le \l \le 0, 
\end{array}\right.$$
$$\Dl_2(\l,K,B,A)=~~~\l K\- (\bea),~\mbox{    for}~\l \ge R_{\bea}.
~~~~~~~~~~~~~~~~$$
See Figure~\ref{pr21d}. Since $\Dl_i(\l,K,B,A)=-H_i(\l,K,A,B)$, 
for $i=1,2$, we get 
\begin{observation} \label{ob_7.2.2}
(a) For $\l>0$ we have
$$\dl_1(B+p,A)\left\{\begin{array}{lll}
\le\l, ~\mbox{for}~p \in \Dl_1(\l,K,B,A);\\
<\l,~\mbox{for}~p \in \Dl_1(\l,\i K,B,A);\\
=\l,~\mbox{for}~p \in \bd\Dl_1(\l,\i K,B,A);\\
\ge\l, ~\mbox{for}~p \in [\Dl_1(\l,\i K,B,A)]^c.\\
%>\l,~\mbox{for}~p \in [\Dl_1(\l,K,B,A)]^c.\\
\end{array}\right.$$
%\dl_1(B+p,A)\left\{\begin{array}{lll}
%\le\l, ~\mbox{for}~p \in \Dl_1(\l,K,B,A);\\
%\ne\l, ~\mbox{for}~p \in \bd[\Dl_1(\l,\i K,B,A)]^c;\\
%\ge\l, ~\mbox{for}~p \in [\Dl_1(\l,\i K,B,A)]^c.\\
%\end{array}\right.$$
(b) For $i=1,2$, $-r_{\bea}\le\l_1\le 0$, and $\l_2\ge R_{\bea}$, 
respectively, we have
$$~~~~\dl_i(B+p,A)\left\{\begin{array}{lll}
\le\l_i, ~\mbox{for}~p \in \Dl_i(\l_i,K,B,A);\\
<\l_i, ~\mbox{for}~p \in \i\Dl_i(\l_i,K,B,A);\\
=\l_i, ~\mbox{for}~p \in \bd\Dl_i(\l_i,K,B,A);\\
\ge\l_i, ~\mbox{for}~p \in \k [\Dl_i(\l_i,K,B,A)]^c.\\
%>\l_i, ~\mbox{for}~p \in [\Dl_i(\l_i,K_i,B,A)]^c.\\
\end{array}\right.$$
%\dl_i(B+p,A)\left\{\begin{array}{lll}
%\le\l_i, ~\mbox{for}~p \in \Dl_i(\l_i,K,B,A);\\
%\ne\l_i, ~\mbox{for}~p \in \bd\Dl_i(\l_i,K,B,A);\\
%\ge\l_i, ~\mbox{for}~p \in \k [\Dl_i(\l_i,K,B,A)]^c.\\
%\end{array}\right.$$
\end{observation}
Thus, $\inf_{p\in R^n}\{\dl_{1(2)}(B+p,A)\}=-r_{\bea}(R_{\bea})$. Hence, we 
can conclude 
\begin{theorem} \label{t4}
$\dl_i(B,A)\sim \Dl_i(\l,K,B,A)$, for $i=1,2$.
\end{theorem}
Note that the Theorem~\ref{t4} also follows from the claim (a) of 
Lemma~\ref{l5}, and from the relationship 
$\Dl_{1(2)}(\l,K,B,A)=\G_{1(2)}(\l,K,\bea)$.
%(Note that the Theorem~\ref{t4} also follows from Lemmas~\ref{l4} and 
%\ref{l5}, and from the relationship $\Dl_i(\l,K,B,A)=\G_i(\l,K,\bea)$, for 
%$i=1,2$.) 
%We also get $\Dl_{1(2)}(\l,K,B,A)=-H_{1(2)}(\l,K,A,B)$.
%---------------------------------------------------------------------------
\begin{figure}[htb]
\begin{center}
\input{pres_21d.pstex_t}
\caption{The objects (a) $\Dl_{1,2}(\l_{1,2},K,B,A)$. Here 
$\dl_{1,2}(B+p,A)=\l_{1,2}$ and $\l_1<0$.} 
\label{pr21d}
\end{center}
\end{figure}
%----------------------------------------------------------------------------
%The objects of $\Dl_{1,2}(\l,K,B,A)$ have the following simple properties: 
%$\Dl_{1(2)}(\l,K,B,A)=-H_{1(2)}(\l,K,A,B)$. The object 
%$\Dl_1(-r_{\bea},K,B,A)$ is the locus of the centers of all largest 
%inscribed balls in $\bea$. The object $\Dl_2(R_{\bea},K,B,A)$ is a singleton 
%point, which is the center of the (unique) smallest enclosing ball of $\bea$. 
%The object $\Dl_2(\l,K,B,A)$ is convex, for any bounded $A$ and $B$, and 
%$\Dl_2(\l,K,B,A)=\Dl_2[\l,K,CH(\bea)]=\Dl_2[\l,K,ext(\bea)]$.

The additional properties of the distances $\dl_{1,2}(B,A)$ and the families 
$\Dl_{1,2}(\l,K,B,A)$ have been studied in \cite{Pk1}.

\subsection*{6 Hausdorff distances and corresponding families of objects}

Since $H(A,B)=\inf\{\l\ge 0\mid B\s \G_1(\l,K,A),A\s \G_1(\l,K,B)\}$ \cite{H}, 
we have $h(B,A)=\inf\{\l\ge 0 \mid B\s \G_1(\l,K,A)\}$. Then $h(B,A)=0$ in 
case where $B\s A$, i.e., the distance $h(B,A)$ does not take into account 
the ``amount'' of containment of $B$ in $A$.

In \cite{Pk3} has been introduced the signed distance 
$\mu(B,A)=\sup_{b\in B}\{\g_1(b,A)\}$, eliminating this shortcoming, and 
it is shown that
$$\mu(B,A)=
\left\{\begin{array}{cc} 
h(B,A),\mbox{ for } B\not \subset A; \\
\eta_1(B,A),\mbox{ otherwise}. \\
\end{array}\right.$$
Then the signed distance $\mu(A,B)$ can be defined as
$$\mu(A,B)=
\left\{\begin{array}{cc} 
h(A,B),\mbox{ for } A\not \subset B; \\
\dl_1(B,A),\mbox{ otherwise}. \\
\end{array}\right.$$
Hence, $H(A,B)=\max\{\mu(A,B),\mu(B,A)\}$.

In \cite{AST} and \cite{Pk3} has been suggested the parametric family of 
objects $M_1(\l,K,B,A)=\G_1(\l,A)\- \cB$, for $\l\ge -r_{\aeb}$, and it is 
shown that $\mu(B,A)\sim M_1(\l,K,B,A)$; see Figure~\ref{pr14}(a). The family 
of objects $M_2(\l,K,B,A)=\G_1(\l,\cB)\- A$, for $\l\ge 0$, has been 
introduced in \cite{AST}, and it is shown (in our notation) that 
$h(A,B)\sim M_2(\l,K,B,A)$, and 
$H(A,B)\sim M_3(\l,K,B,A)=[ M_1(\l,K,B,A)\cap M_2(\l,K,B,A)]$, respectively. 
See Figure~\ref{pr14}(b).
% (The objects of $M_{1-3}(\l,K,B,A)$ have used in 
%\cite{AST} for computing the minimum Hausdorff distance.)  
Clearly, for $-r_{\bea}\le\l\le 0$, we have $\mu(A,B)\sim M_2(\l,K,B,A)$.

For notational convenience, the distances $\mu(B,A)$, $\mu(A,B)$, and $H(A,B)$ 
are sometimes denoted by $m_1(B,A)$, $m_2(B,A)$, and $m_3(B,A)$, respectively. 
Then we get
%---------------------------------------------------------------------------
\begin{figure}[htb]
\begin{center}
\input{pres_14.pstex_t}
\caption{The objects (a) $M_{1,2}(\l_{1,2},K,B,A)$. Here, respectively, 
$\mu(B+p,A)=\l_1>0$, $\mu(A,B+p)=\l_2>0$, and  
%$\i A\ominus\i\cB=\i (A\ominus\cB)$ and $\i\cB\ominus\i A=\i(\cB\ominus A)$, 
$\bd\G_1(\l_1,K,A)=\bd\G_1(\l_1,\i K,\i A)$, 
$\bd\G_1(\l_2,K,\cB)=\bd\G_1(\l_2,\i K,\i \cB)$. 
} 
\label{pr14}
\end{center}
\end{figure}
%----------------------------------------------------------------------------
\begin{theorem} \label{t3}
$m_i(B,A)\sim M_i(\l,K,B,A)$, for $i=1 - 3$.
\end{theorem}
The objects $M_{1-3}(\l,K,B,A)$ have the following simple properties:
$$M_{1(2)}(\l,K,B,A)=-M_{2(1)}(\l,K,A,B);~M_3(\l,K,B,A)=-M_3(\l,K,A,B).$$
%We turn to studying the properties of Hausdorff distances and their 
%corresponding families.
%-------------------------------------------------------------------------
\begin{observation} \label{ob_8.1}
(a) For $i=1 - 3$, and $\l>0$ we have
$$m_i(B+p,A)\left\{\begin{array}{lll}
\le\l,~\mbox{for}~p \in M_i(\l,K,B,A);\\
<\l,~\mbox{for}~p \in \i M_i(\l,\i K,\i B,\i A);\\
=\l,~\mbox{for}~p \in M_i(\l,K,B,A)\bsl\i M_i(\l,\i K,\i B,\i A);\\
\ge\l,~\mbox{for}~p \in \k [M_1(\l,\i K,\i B,\i A)]^c.\\
%>\l,~\mbox{for}~p \in [M_i(\l,K,B,A)]^c.\\
\end{array}\right.$$
%$$~~~~~~m_i(B+p,A)\left\{\begin{array}{lll}
%\le\l,~\mbox{for}~p \in M_i(\l,K,B,A);\\
%\ne\l,~\mbox{for}~p \in [M_i(\l,K,B,A)]^c\cup \i M_i(\l,\i K,\i B,\i A);\\
%\ge\l,~\mbox{for}~p \in \k [M_1(\l,\i K,\i B,\i A)]^c.\\
%\end{array}\right.$$
(b) For $i=1,2$, $-r_{\aeb}\le\l_1\le 0$, and $-r_{\bea}\le\l_2\le 0$ we have
%$$\mu(B+p,A)\left\{\begin{array}{lll}
%<\l, ~\mbox{for}~p \in \i M_1(\l,K,B,A);\\
%=\l, ~\mbox{for}~p \in \bd M_1(\l,K,B,A);\\
%>\l, ~\mbox{for}~p \in [M_1(\l,K,B,A)]^c.\\
%\end{array}\right.~~~~~~~~~~~~~~~~~~~~~~~~~~~~~~~$$
$$m_i(B+p,A)\left\{\begin{array}{lll}
\le\l_i, ~\mbox{for}~p \in M_i(\l_i,K,B,A);\\
<\l_i, ~\mbox{for}~p \in \i M_i(\l_i,K,B,A);\\
=\l_i, ~\mbox{for}~p \in \bd M_i(\l_i,K,B,A);\\
\ge\l_i, ~\mbox{for}~p \in \k[M_i(\l_i,K,B,A)]^c.\\
%>\l_i, ~\mbox{for}~p \in [M_i(\l_i,K,B,A)]^c.\\
\end{array}\right.~~~~~~~~~~~~~~~~~~~~$$
%m_i(B+p,A)\left\{\begin{array}{lll}
%\le\l_i, ~\mbox{for}~p \in M_i(\l_i,K,B,A);\\
%\ne\l_i, ~\mbox{for}~p \in [\bd M_i(\l_i,K,B,A)]^c;\\
%\ge\l_i, ~\mbox{for}~p \in \k[M_i(\l_i,K,B,A)]^c.\\
%\end{array}\right.$$
\end{observation}

{\bf Remark 3} In \cite{Pk1} it is shown that, in contrast to the distances 
considered in Sections 2 -- 5, the region where the distance $\mu(B+p,A)$ 
(resp., $\mu(A,B+p)$) is equal to $\l$ may have the non-empty interior.

The additional properties of the Hausdorff distances and their corresponding 
families have been studied in \cite{Pk1}.

\subsection*{7 Translational distances between geometric objects and 
translational geometric situations}

The distances considered in Sections 2 -- 6 are referred to as the 
{\em translational} distances (or $T$-distances). Let  
$\CT\D(B,A)=\{\g_{1,2}(B,A),\eta_{1,2}(B,A),\dl_{1,2}(B,A),
m_{1-3}(B,A)\}$ be a collection of the $T$-distances between $B$ and $A$, and 
let $\o(B,A)\in \CT\D(B,A)$. The $T$-distances $\o(B,A)$ and their 
corresponding families $\O(\l,K,B,A)$ have the following properties: 
$$\o(B+p,A)=\o(B,A-p)=\o[B+\a\cdot p,A-(1-\a)\cdot p],
~\mbox{for}~0\le \a \le 1;$$
$$\o(B^\t\pm p,A^\t\pm q)=\o(B^\t\mp q,A^\t\mp p);$$
$$\O(\l_1,K,B,A)\se\O(\l_2,K,B,A),~\mbox{for any bounded $A$ and $B$, and}~
\l_1\le\l_2;$$ 
$$\O[\l,K,B+\a\cdot p,A-(1-\a)\cdot p]
=\O(\l,K,B,A)-p,~\mbox{for}~0\le \a\le 1;$$ 
$$\O(\l,K,B^\t\pm p,A^\t\pm q)=\O(\l,K,B^\t\mp q,A^\t\mp p)=
[\O(\l,K,B,A)]^\t\mp p \pm q.$$
From the last relationship it follows that $\o(B\pm p,A\pm q)\le\l$, for 
$\pm p\mp q\in\O(\l,K,B,A)$.

%\begin{equation}
%\o(B\pm p,A\pm q)\left\{\begin{array}{lll} 
%\le\l, ~\mbox{for}~\pm p \mp q \in \O(\l,K,B,A);\\
%>\l, ~\mbox{for}~\pm p \mp q \in  [\O(\l,K,B,A)]^c.\\
%\end{array}\right.
%\label{eq_9.1}
%\end{equation}
%By properties of Minkowski operations, it can easily be shown that
%$$\G_2(\l,K,B,A)\se H_2(\l,K,B,A)\se H_1(\l,K,B,A)\se M_1(\l,K,B,A)\se 
%\G_1(\l,K,B,A),~\mbox{for}~\aeb\ne\es;$$
%$$\G_2(\l,K,B,A)\se \Dl_2(\l,K,B,A)\se \Dl_1(\l,K,B,A)\se M_2(\l,K,B,A)\se 
%\G_1(\l,K,B,A),~\mbox{for}~\bea\ne\es.$$
%$$\G_2(\l,K,B,A)\se H_2(\l,K,B,A)\se H_1(\l,K,B,A)\se M_1(\l,K,B,A)\se 
%\G_1(\l,K,B,A),$$
%for $\aeb\ne\es$, and
%$$\G_2(\l,K,B,A)\se \Dl_2(\l,K,B,A)\se \Dl_1(\l,K,B,A)\se M_2(\l,K,B,A)\se 
%\G_1(\l,K,B,A),$$
%for $\bea\ne\es$, respectively. (An equalities hold in case where $A$ and $B$ 
%are a singleton points.)

%Analysis of the relative position of an object $B$ with respect to an object 
%$A$ consists in computing the {\it signes} and/or {\it values} of the 
%$T$-distances between $B$ and $A$. For instance,

By the previous observations, we get
$$\begin{array}{lll} 
(A \cap B = \es)\wedge(B\not\subset A) 
& \iff \big[\g_1(B,A)>0\big]\bigwedge
\Big\{\big[\eta_1(B,A)>0\big]\bigvee\big[\mu(B,A)>0\big]\Big\}; \\
%(A \cap B \ne \es)\wedge(B\s A)        
%& \iff \big[\g_1(B,A)\le 0\big]\bigwedge
%\Big\{\big[\eta_1(B,A)\le 0\big]\bigvee\big[\mu(B,A)\le 0\big]\Big\}; \\
(A \cap B \ne \es)\wedge(A\s B)         
& \iff \big[\g_1(B,A)\le0\big]\bigwedge
\Big\{\big[\dl_1(B,A)\le 0\big]\bigvee\big[\mu(A,B)\le 0\big]\Big\}. \\
\end{array}$$
Hence, the various situations of the relative position of objects $A$ and 
$B$ can be described by the system of constraints on the $T$-distances between 
$A$ and $B$.
%It is clear that the inequalities in the above relationships describe a 
%various situations of the relative position of objects $A$ and $B$.
\begin{definition} \label{d20} The relationship 
$\nu(B+p,A)=[\o(B+p,A) \odot \l]$, where $\odot \in \{<,=,>\}$, is called the 
{\it primitive translational geometric situation} (PTGS) of an object $B$ with 
respect to an object $A$.
\end{definition}
%For example, $\nu_1(B+p,A)=[\g_1(B+p,A)=\l_1]$ and 
%$\nu_2(B+p,A)=[\dl_2(B+p,A)>\l_2]$ are the primitive TGS's of $B$ relative to 
%$A$.

Let us consider the PTGS $\nu(B+p,A)$ as an event. Then the class of all the 
possible PTGS's $\S^+(B,A)$ forms an {\it algebra} $\A_{\CT}$ of events 
\cite{KK}. For $\S^+(B,A)$ permitting the following definitions:

1. The {\it union} of events: $\nu_1(B+p,A)\vee\nu_2(B+p,A)$.

2. The {\it intersection} of events: $\nu_1(B+p,A)\wedge\nu_2(B+p,A)$.

3. The {\it complement} of event: $[\nu(B+p,A)]^c$.

4. The {\it certain} event $I$ is the union of all the PTGS's in $\S^+(B,A)$.

5. The {\it impossible} event $0$ is an impossible relative position of 
objects, for given $\nu(B+p,A)$.

The class $\S(B,A)$ of PTGS's, comprising $\S^+(B,A)$ and $0$, forms a 
completely additive Boolean algebra; see \cite{KK}.

Let us consider the union of PTGS's. (Recall that $\o(B,A)\in\CT\D(B,A)$.) 
Then we get
%$$[\o(B+p,A)\le\l]=[\o(B+p,A)<\l]\vee [\o(B+p,A)=\l];$$
%$$[\o(B+p,A)\ne\l]=[\o(B+p,A)<\l]\vee [\o(B+p,A)>\l];$$
%$$[\o(B+p,A)\ge\l]=[\o(B+p,A)>\l]\vee [\o(B+p,A)=\l].$$
$$[\o(B+p,A)\le\l]=[\o(B+p,A)<\l]\vee [\o(B+p,A)=\l];$$
\begin{equation}
[\o(B+p,A)\ne\l]=[\o(B+p,A)<\l]\vee [\o(B+p,A)>\l];
\label{eq_9.2}
\end{equation}
$$[\o(B+p,A)\ge\l]=[\o(B+p,A)>\l]\vee [\o(B+p,A)=\l].$$
(Clearly, $[\o(B+p,A)<\l]\wedge [\o(B+p,A)>\l]=0$.) We define also the TGS of 
type $[\o(B+p,A)\to \min]$, since in Sections 4 -- 6 have been obtained 
the possible minimal values of the $T$-distances. We next add these TGS's to 
the set of PTGS's: 
%-----------------------------------------------------------------------
\begin{definition} \label{d21} The relationship 
$\nu(B+p,A)=[\o(B+p,A) \odot \l\mbox{ or }\o(B+p,A)\to \min]$, where 
$\odot \in \{<,\le,=,\ne,\ge,>\}$, is called the {\it basic translational 
geometric situation} (BTGS) of $B$ relative to $A$.
\end{definition}
%For instance, $\nu_1(B+p,A)=[\g_1(B+p,A)\le\l]$ and 
%$\nu_2(B+p,A)=[\g_2(B+p,A)\to \min]$ are the BTGS's. Note that 
Since
$[\o(B+p,A)<\l]^c=[\o(B+p,A)\ge\l]$, $[\o(B+p,A)=\l]^c=[\o(B+p,A)\ne\l]$, and 
$[\o(B+p,A)>\l]^c=[\o(B+p,A)\le\l]$, then, by the relationships of 
(\ref{eq_9.2}), 
and, by DeMorgan's laws, we obtain the 
following useful relationships:
$$[\o(B+p,A)<\l]=[\o(B+p,A)\le\l]\wedge [\o(B+p,A)\ne\l];$$
$$[\o(B+p,A)=\l]=[\o(B+p,A)\le\l]\wedge [\o(B+p,A)\ge\l];$$
$$[\o(B+p,A)>\l]=[\o(B+p,A)\ge\l]\wedge [\o(B+p,A)\ne\l].$$

\begin{definition} \label{d30}The Boolean function 
$\nu(B+p,A)=f[\nu_1(B+p,A),\ldots,\nu_k(B+p,A)]$ of BTGS's 
$\nu_i(B+p,A)$, for $i=1,\ldots,k$, is called the {\it translational geometric 
situation} (TGS) of $B$ with respect to $A$.
\end{definition}
A particular TGS describes the constraints on the relative position of 
objects, and can be used as the {\it objective function} in the spatial 
planning problems.
% In case where $p=O$, the TGS can be interpreted as the 
%generalized Boolean {\it distance query}; see \cite{LinM} for detailes.
%For instance, in 
%case where $\nu_1(B+p,A)=[\g_1(B+p,A)\ge\l_1]$, 
%$\nu_2(B+p,A)=[\g_2(B+p,A)\le\l_2]$, and 
%$\nu(B+p,A)=\nu_1(B+p,A)\wedge\nu_2(B+p,A)$ we obtain the TGS
%$$\nu(B+p,A)=[\g_1(B+p,A)\ge\l_1]\wedge[\g_2(B+p,A)\le\l_2].$$

\subsection*{8 Constructing the feasible region of an object for given 
translational geometric situation}

%Let us consider a correspondence relation $\o(B,A)\sim \O(\l,K,B,A)$, and a 
%collection $\C(B,A)$ of point sets
%$$\C(B,A)=\bigg\{\i \O(\l,B,A),\O(\l,B,A),\bd \O(\l,B,A),
%[\bd \O(\l,B,A)]^c,[\O(\l,B,A)]^c,\k [\O(\l,B,A)]^c\bigg\}.$$
Let $\O(\l,K,B,A)$ is the corresponding family of the distance $\o(B,A)$. From 
the observations of Sections 4 -- 6, and from the Definition \ref{d21} it 
follows that, for given BTGS $\nu(B+p,A)$, all the feasible translations $B+p$ 
of $B$ with respect to $A$ are obtained if and only if $p$ belongs to the 
region $N(B,A)$, where $\nu(B+p,A)$ is true. For example, 
$$\begin{array}{lll}
\nu_1(B+p,A)=[\o(B+p,A)\le\l_1] & \iff p\in N_1(B,A)=\O(\l_1,K,B,A); \\
\nu_2(B+p,A)=[\o(B+p,A)>\l_2] &\iff p\in  N_2(B,A)=[\O(\l_2,K,B,A)]^c. \\
\end{array}$$
The correspondence between $\nu(B+p,A)$ and $N(B,A)$ is denoted by 
$\nu(B+p,A)\sim N(B,A)$.
%Thus, we get the following:
%\begin{observation} \label{f2} Let $\nu(B+p,A)$ be the BTGS, and 
%$\nu(B+p,A)\sim N(B,A)$. Then 
%$$\nu(B\pm p,A\pm q)=
%\Big\{\g_1\Big[\pm p \mp q,N(B,A)\Big]\le 0\Big\}.$$
%\end{observation}

Consider the TGS $\nu(B+p,A)=\nu_1(B+p,A)\wedge\nu_2(B+p,A)$, where the  
BTGS's $\nu_{1,2}(B+p,A)$ are as above. Clearly, its corresponding region is 
$N(B,A)=N_1(B,A)\cap N_2(B,A)$. In case where $\l_1\le \l_2$, we have 
$N(B,A)=\es$, and therefore $\nu(B+p,A)$ is an impossible TGS, i.e., 
$\nu(B+p,A)=0$. For the TGS $\nu(B+p,A)=\nu_1(B+p,A)\vee\nu_2(B+p,A)$ we have 
$N(B,A)=N_1(B,A)\cup N_2(B,A)$. If $\l_1\ge\l_2$ then $N(B,A)=R^n$. (Note that 
in this case $\nu(B+p,A)\ne I$.)

Consider next the TGS $\nu(B+p,A)=[\o(B+p,A)\le\l_1]\vee[\o(B+p,A)\le\l_2]$, 
where $\l_1\le\l_2$. Then $N(B,A)=\O(\l_1,K,B,A)\cup\O(\l_2,K,B,A)$ and 
$\O(\l_1,K,B,A)\se\O(\l_2,K,B,A)$, for $\l_1\le\l_2$. Hence, 
$N(B,A)=\O(\l_2,K,B,A)$. That is, the TGS $\nu(B+p,A)$ corresponding to the 
region $N(B,A)$ can be represented in different ways, e.g., as 
$\nu(B+p,A)=[\o(B+p,A)\le\l_2]$ or 
$\nu(B+p,A)=\bigvee_{i=1}^n[\o(B+p,A)\le\l_i]$, where $\l_i\le\l_2$, for 
$i=1,\dots,n$. Thus, in general, the TGS $\nu(B+p,A)$ does not unique relative 
to its corresponding region $N(B,A)$. However, for given $N(B,A)$, the unique 
(minimal) corresponding TGS $\nu(B+p,A)$ can be constructed. Finally, if in 
the above example, we let $N(B,A)=\O(\l,K,B,A)\cup\O(\l_2,K,B,A)$, where 
$\l_1<\l\le\l_2$, then we obtain that the TGS $[\o(B+p,A)\le\l_1]$ does not 
corresponds to the region $\O(\l,K,B,A)$, however $\nu(B+p,A)\sim N(B,A)$ in 
this case. Therefore, we get the following:
\begin{proposition} \label{p1} (a) The TGS 
$\nu(B+p,A)=f[\nu_1(B+p,A),\ldots,\nu_k(B+p,A)]$ corresponds to the region 
$N(B,A)=F[N_1(B,A),\ldots,N_k(B,A)]$ if (but not only if) 
$\nu_i(B+p,A)\sim N_i(B,A)$, for $i=1,\ldots, k$, and the operations of union, 
intersection, and complement of the BTGS's $\nu_i(B+p,A)$ of $\nu(B+p,A)$ 
correspond to the operations of union, intersection, and complement of regions 
$N_i(B,A)$ of $N(B,A)$. (b) $\nu(B+p,A)$ is an impossible TGS if and only if 
$N(B,A)=\es$. (c) $\nu(B+p,A)$ is a certain TGS only if (but not if) 
$N(B,A)=R^n$.
\end{proposition} 
The region $N(B,A)$ is called the {\it solution} of the TGS $\nu(B+p,A)$. 
(Note that the region $N_i(B,A)$ of $N(B,A)$ is solution of the BTGS 
$\nu_i(B+p,A)$, for $i=1,\ldots, k$.)
%-----------------------------------------------------------------------
%{\bf Remark 4} Let $\nu(B+p,A)$ be the BTGS concerning the $T$-distance 
%$\o(B+p,A)$, and let $\o(B+p,A)\sim\O(\l,K,B,A)$, and 
%$\nu(B+p,A)\sim N(B,A)$, respectively. In the next paragraph we assume, for 
%simplicity, that the region 
%$N(B,A)$ does not have coincident faces/edges and/or isolated points, which 
%are removed from the interior of the objects of $\O(\l,K,B,A)$, for any $\l$. 
%Then, for the corresponding objects of the $T$-distances $\g_1(B,A)$, 
%$\eta_1(B,A)$, and $\dl_1(B,A)$, the following relationships hold:
%$$\G_1(\l,K,B,\i A)=\i \G_1(\l,K,B,A),~\mbox{for}~\l\ge 0;
%~~\adbi=\i(\adb),~\mbox{for}~\l\le 0;$$
%$$H_1(\l,\i K,B,A)=\i H_1(\l,K,B,A),~\mbox{for}~\l>0;~~
%\Dl_1(\l,\i K,B,A)=\i \Dl_1(\l,K,B,A),~\mbox{for}~\l>0.$$
%For objects corresponding to the Hausdorff distances we have
%$$M_{1-3}(\l,\i K,\i B,\i A)=M_{1-3}(\l,K,B,A),~\mbox{for}~\l>0.$$
%The last relationship holds in case where $\G_1(\l,K,\i A)=\i \G_1(\l,K,A)$, 
%and $\G_1(\l,K,\i \cB)=\i \G_1(\l,K,\cB)$, respectively. Thus, in general, we 
%get
%\begin{equation}

{\bf Remark 4} Let $\nu(B+p,A)$ be the BTGS concerning the $T$-distance 
$\o(B+p,A)$, and let $\o(B+p,A)\sim\O(\l,K,B,A)$, and $\nu(B+p,A)\sim N(B,A)$, 
respectively. In the next paragraph we assume, for simplicity, that the region 
$N(B,A)$ does not have coincident faces/edges and/or isolated points, which 
are removed from the interiors of the objects of $\O(\l,K,B,A)$, for any $\l$. 
Then we get
$$\o(B+p,A)\left\{\begin{array}{lll} 
\le\l,~\mbox{for}~p \in \O(\l,K,B,A);\\
<\l,~\mbox{for}~p \in \i \O(\l,K,B,A);\\
=\l,~\mbox{for}~p \in \bd \O(\l,K,B,A);\\
%>\l, ~\mbox{for}~p \in  [\O(\l,K,B,A)]^c.\\
\ge\l, ~\mbox{for}~p \in \k [\O(\l,K,B,A)]^c.
\end{array}\right.$$
%\o(B+p,A)\left\{\begin{array}{lll} 
%\le\l, ~\mbox{for}~p \in \O(\l,K,B,A);\\
%\ne\l, ~\mbox{for}~p \in [\bd \O(\l,K,B,A)]^c;\\
%\ge\l, ~\mbox{for}~p \in \k [\O(\l,K,B,A)]^c.\\
%\end{array}\right.$$
%\label{eq_10.1}
%\end{equation}
In this special case the set $\bd \O(\l,K,B,A)$ is the surface of the 
function $\o(B+p,A)$.

The types of objects satisfying the above assumption are sufficiently wide. 
For example, the convex objects, the polygons/polytops and/or curved objects 
in general position, i.e., which do not have parallel edges/faces. Thus, one 
can to construct the solution of a particular TGS according to the more 
simpler relationships than that is done in Sections 4 -- 6. See \cite{Pk1} for 
details.
%Observation~\ref{ob_6.2.1}, and in the claim (a) of 
%Observations~\ref{ob_7.1.1}, \ref{ob_7.2.2}, and \ref{ob_8.1}, respectively.
%Clearly,
%\begin{equation}
%\o(B+p,A)\left\{\begin{array}{lll} 
%\le\l, ~\mbox{for}~p \in \O(\l,K,B,A);\\
%\ne\l, ~\mbox{for}~p \in [\bd \O(\l,K,B,A)]^c;\\
%\ge\l, ~\mbox{for}~p \in \k [\O(\l,K,B,A)]^c.\\
%\end{array}\right.
%\label{eq_10.2}
%\end{equation}
%Note that the above relationships hold for objects of 
%$\G_1(\l,K,B,A)$, $H_1(\l,K,B,A)$, and $\Dl_1(\l,K,B,A)$, in case where 
%$\l\le 0$, for any $A$, and any bounded $B$, and for objects of 
%$\G_2(\l,K,B,A)$, $H_2(\l,K,B,A)$, and $\Dl_2(\l,K,B,A)$, for any bounded $A$ 
%and $B$.
%------------------------------------------------------------------------
\paragraph{Examples.} Consider the TGS's $\nu_l(B+p,A)$ and their solutions 
$N_l(B,A)$, denoted by $\nu_l$, $N_l$, for short. Denote also $\o(B+p,A)$, 
$\O(\l,K,B,A)$ by $\o$, $\O(\l)$. Below we let $i(j)=1,2$.
$${\bf 1}~~~ \nu_1=(\g_i\le\l_i) \wedge (\g_j\ge\l_j); ~~~
N_1=\G_i(\l_i)\cap\k[\G_j(\l_j)]^c=\G_i(\l_i)\bsl\i\G_j(\l_j).$$
$${\bf 1a}~~~ \nu_{1a}=(\g_i<\l_i) \wedge (\g_j>\l_j); ~~~
N_{1a}=\i\G_i(\l_i)\cap[\G_j(\l_j)]^c=\i\G_i(\l_i)\bsl\G_j(\l_j).$$
$${\bf 1b}~~~ \nu_{1b}=(\g_i\le\l_i) \wedge (\g_j>\l_j); ~~~
N_{1b}=\G_i(\l_i)\cap[\G_j(\l_j)]^c=\G_i(\l_i)\bsl\G_j(\l_j).$$
$${\bf 1c}~~~ \nu_{1c}=(\g_i\ge\l_i) \wedge (\g_j\ge\l_j); ~~~
N_{1c}=\k[\G_i(\l_i)]^c\cap\k[\G_j(\l_j)]^c=\k[\G_i(\l_i)\cup\G_j(\l_j)]^c.$$
$${\bf 2}~~~ \nu_2=(\g_i=\l_i) \wedge (\g_j=\l_j); ~~~
N_2=\bd\G_i(\l_i)\cap\bd\G_j(\l_j).$$
$${\bf 3}~~~ \nu_3=(\g_1\ge\l_1) \wedge (\g_2 \to \min),
\mbox{ where }\l_1\ne r_{(\adb)};$$
$$N_3=\k[\G_1(\l_1)]^c\cap\G_2(R_{\adb})=\G_2(R_{\adb})\bsl\i\G_1(\l_1).$$
$${\bf 4}~~~\nu_4=(\eta_1 \to \min); ~~~N_4=H_1(-r_{\aeb}).$$
$${\bf 5}~~~ \nu_5=(\eta_1\le\l_1\le 0)\wedge(\eta_2\to \min); ~~~
N_5=H_1(\l_1)\cap H_2(R_{\aeb}).$$
$${\bf 6}~~~ \nu_6=[(\g_1=\l_1) \wedge (\g_2=\l_{\min})]\ne 0; ~~~
N_6=\bd\G_1(\l_1)\cap\bd\G_2(\l_{\min}),$$
$$\mbox{where }\l_{\min}=\left\{\begin{array}{cc} 
\inf\{\l\mid \G_2(\l)\dot \s\G_1(\l_1)\},
~\mbox{for}~\G_2(R_{\adb})\in \G_1(\l_1); \\
\inf\{\l\mid \G_2(\l)\dot \cap\G_1(\l_1)\},
~\mbox{for}~\G_2(R_{\adb})\notin \G_1(\l_1). \\
\end{array}\right.$$
%$${\bf 7}~~~ \nu_7=[(\g_1=\l_1) \wedge (\g_2=\l_{\max})]\ne 0; ~~~
%N_7=\bd\G_1(\l_1)\cap\bd\G_2(\l_{\max}),$$
%$$\mbox{where }\l_{\max}= \inf\{\l\mid \G_1(\l)\dot \s\G_2(\l)\}.$$
%See Figures~\ref{pr16b} and \ref{pr16a}.
See Figure \ref{pr16a}. The analysis of the Examples 1 -- 6 can be found 
in \cite{Pk1}.
%---------------------------------------------------------------------------
%\begin{figure}[htb]
%\begin{center}
%\input{pres_16b.pstex_t}
%\caption{The regions $N_l$ corresponding to the TGS's $\nu_l$, for 
%$l=1,\ldots,6$, in case where $\G_2(R_{\adb})\in \G_1(\l_1)$. Here $A$ and 
%$B$ are quadrangles, $i=2$, $j=1$, and $\l_1<\l_2$, respectively. The region 
%$N_1$ is four-connected, $N_2=\{p_1,\ldots,p_8\}$, and $N_3=\es$, 
%respectively. The regions $N_4$ and $N_5$ are singleton points; 
%$N_6=\{q_1,\ldots,q_4\}$.}
%\label{pr16b}
%\end{center}
%\end{figure}
%----------------------------------------------------------------------------
%---------------------------------------------------------------------------
\begin{figure}[htb]
\begin{center}
\input{pres_16a.pstex_t}
\caption{The corresponding regions $N_l$ of TGS's $\nu_l$, for $l=1,\ldots,6$, 
in case where $\G_2(R_{\adb})\not\in\G_1(\l_1)$. Here $i=2$, $j=1$, and 
$\l_1<\l_2$, respectively. The region $N_1$ is simply connected, 
$N_2=\{p_1,p_2\}$, and $N_3=\G_2(R_{\adb})$, respectively. The region $N_4$ is 
a curve, $N_5=\es$, and $N_6=\{q_1,q_2,q_3\}$.}
\label{pr16a}
\end{center}
\end{figure}
%----------------------------------------------------------------------------
%\input{pres_161.pstex_t}
%\caption{The region $N_6$ corresponding to TGS $\nu_6$. Here the region 
%$N_6(B,A)$ consists of the four singleton points.}
%Note that $N_{1a}=\i N_1$, and that in the Examples 1--1b in case where 
%$\G_i(\l_i) \s \G_j(\l_j)$, we have $N_{1-1b}=\es$ and $\nu_{1-1b}=0$. In the 
%Example 2, if  $\bd\G_i(\l_i) \s \bd\G_j(\l_j)$ 
%then $N_2=\es$, i.e., $\nu_2=0$ in this special case. In Examples 3 and 6 we 
%consider two special cases. If $\G_2(R_{\adb})\in \G_1(\l_1)$ then  
%$N_3=\es$, 
%$\nu_3=0$, and in Example 6 the value of $\l_{\min}$ corresponds to the inner 
%touching of $\G_2(\l_{\min})$ relative to $\G_1(\l_1)$. (See 
%Figure~\ref{pr16b}.) In case where $\G_2(R_{\adb})\notin \G_1(\l_1)$ we have 
%$N_3=\G_2(R_{\adb})$, and therefore $\nu_3=(\g_2 \to \min)\ne 0$, and in 
%Example 6 the value of $\l_{\min}$ corresponds to the outer touching of 
%$\G_2(\l_{\min})$ relative to $\G_1(\l_1)$. (See Figure~\ref{pr16a}.) 
%Analogously, in Example 5 we have $\nu_5=(\eta_2\to \min)\ne 0$ and 
%$N_5=H_2(R_{\aeb})$, for $H_2(R_{\aeb})\in H_1(\l_1)$, and $\nu_5=0$, 
%$N_5=\es$, otherwise, respectively. (See Figures~\ref{pr16b} and \ref{pr16a}.)

{\bf Remark 5} The region $N(B,A)$ may have various topology. It can be open 
or closed, regular or non-regular, bounded or unbounded, connected or 
disconnected. So, in the above examples the region $N_1$ is closed bounded, 
whereas the region $N_{1a}$ is open bounded; the region $N_{1c}$ is closed 
unbounded. In case where $N(B,A)$ contains the subset of its boundary, it is 
neither open nor closed. Thus, in general, $N(B,A)$ is an object with 
{\it non-manifold} boundary. The region $N_{1b}$ give an example of such an 
object. In case where $i=j$, $\l_i>\l_j$, and $\l_{i,j}>0$, respectively, 
the object $N_1$ is regular, i.e.,$N_1=\G_i(\l_i)\bsl^*\G_j(\l_j)$. (Note that 
in case where $\l_i=\l_j$ we get $N_1=\bd\G_i(\l_i)$, i.e., the object $N_1$ 
is non-regular.)

\begin{figure}[htb]
\begin{center}
\input{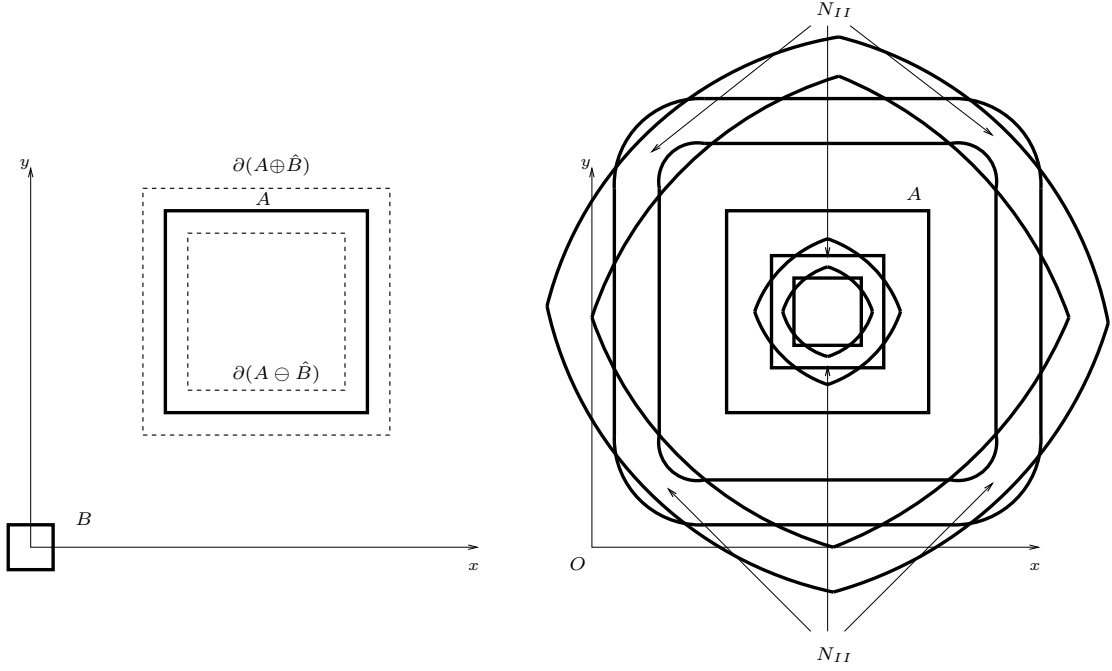}
\caption{Illustration for the Problem II. Here the region $N_{II}$ is 
5-connected.} 
\label{pr19a}
\end{center}
\end{figure}

Let us next consider the following problem:

{\bf Problem II} Find the region $N_{II}$, corresponding to the TGS
$$\nu_{II}=\big[(\l_1\le \g_1\le\l_2)\wedge(\l_3\le \g_2\le\l_4)\big] 
\vee\big[(\l_5\le \eta_1\le\l_6)\wedge(\l_7\le \eta_2\le\l_8)\big].$$
By Example 1, the solution is the region
$$N_{II}=\Big\{\big[\G_1(\l_2) \bsl\i\G_1(\l_1)\big]\cap 
\big[\G_2(\l_4) \bsl\i\G_2(\l_3)\big]\Big\}
\cup\Big\{\big[H_1(\l_6) \bsl\i H_1(\l_5)\big]\cap 
\big[H_2(\l_8) \bsl\i H_2(\l_7)\big]\Big\}.$$
See Figure~\ref{pr19a}. Note that in case where $\l_1>\l_2$ or $\l_3>\l_4$, 
and $\l_5>\l_6$ or $\l_7>\l_8$ we have $N_{II}=\es$ and $\nu_{II}=0$. 
%---------------------------------------------------------------------------
%----------------------------------------------------------------------------
%\input{pres_4.pstex_t} 
%\caption{A region $N_3(B,A)$ in the special cases, where it is a simply 
%connected region. Here (a) $N_3(B,A)=\G_2(\l_4,B,A) \bsl^* \G_2(\l_3,B,A)$, 
%and (b) $N_3(B,A)=\G_1(\l_2,B,A) \bsl^* \G_1(\l_1,B,A)$.}

Let us turn to solve the problem, formulated in subsection 1.3. The following 
regions $N_I'$, for $\odot_1=\wedge$, $\odot_2=\vee$, and $N_{I}''$, for 
$\odot_1=\vee$, $\odot_2=\wedge$, are the solutions of the Problem I: 
$$N_I'=\big[\G_1(\l_1)\bsl\i {\G_1}(\l_2)\big]\cup
\big[\G_2(\l_3) \bsl\i {\G_2}(\l_4)\big] \\
\cup\big[H_1(\l_5)\bsl\i H_1(\l_6)\big];$$
$$N_I''=\big[\G_1(\l_1) \cup \k[\G_1(\l_2)]^c\big]\cap
[\G_2(\l_3)\cup\k[\G_2(\l_4)]^c\big] \\
\cap\big[H_1(\l_5) \cup \k[H_1(\l_6)]^c\big].$$
Clearly, if $\l_1<\l_2$, $\l_3<\l_4$, and $\l_5<\l_6$, then $N_{I}'=\es$ 
and $\nu_{I}'=0$. In case where $\l_1\ge\l_2$, $\l_3\ge\l_4$, and 
$\l_5\ge\l_6$ we have  $N_{I}''=R^d$. (However, $\nu_{I}''\ne I$ in this case.)
%-----------------------------------------------------------------------

%Let $\A=\{A_1,\ldots,A_n\}$ be a collection of objects of $R^n$, and 
%$\nu_i(B+p,A_i)$ and $N_i(B,A_i)$ be the TGS of an object $B$ relative to 
%$A_i$ and its solution, respectively, for $i=1,\ldots,n$. Then we define the 
%{\it complex} TGS $\nu(B+p,\A)=f[\nu_1(B+p,A_1),\ldots,\nu_n(B+p,A_n)]$, of 
%$B$ with respect to $\A$, and, by Proposition {\ref{p1}}, obtain its solution 
%as $N(B,\A)=F[N_1(B,A_1),\ldots,N_n(B,A_n)]$. Then the following useful 
%relationships hold:
%\begin{equation}
%\Big[\bigvee_{i=1}^n\nu_i(B+p,A_i)\Big]^c
%=\bigwedge_{i=1}^n\Big[\nu_i(B+p,A_i)\Big]^c;~~
%\Big[\bigwedge_{i=1}^n\nu_i(B+p,A_i)\Big]^c
%=\bigvee_{i=1}^n\Big[\nu_i(B+p,A_i)\Big]^c.
%\label{eq_10.1}
%\end{equation}

The additional properties of the TGS's have been studied in \cite{Pk1}.

\subsection*{9 Applications}

In this section we consider the spatial planning problems with more general 
and more complex constraints on the distances between geometric objects. 
%In other words, we consider the spatial planning problem, for given 
%translational geometric situation. 
We also briefly consider the several other types of geometric situations: the 
translational geometric situation in a given direction, the rotational, and  
the dynamic geometric situations, respectively.

\subsubsection*{9.1 Findspace problem}

\begin{figure}[htb]
\begin{center}
\input{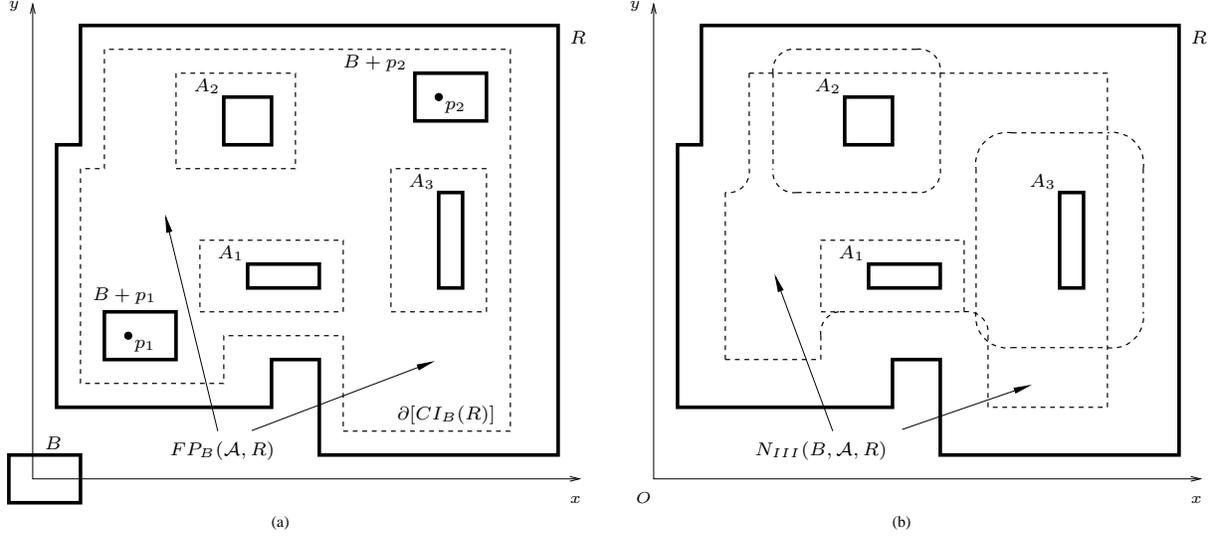}
\caption{The translational Findspace problem in case where 
$\A=\{A_1,A_2,A_3\}$. (a) Here the region $FP_B(\A,R)$ is connected, and 
$p_{1,2}$ are the free positions of $B$. Dashed lines show $\bd [CO_B(\A)]$ 
and $\bd [CI_B(R)]$, respectively. (b) Illustration for the Problem III. Here 
$N_{III}(B,\A,R)$ is a two-connected region and $\l_1=0$, $\l_2=|\l_R|$, and 
$\l_3=1.5 \cdot\l_2$, respectively. 
Dashed curves show $\bd [CO_B(\l_{\A},\A)]$ and $\bd [CI_B(\l_R,R)]$.
}
\label{pr14dd}
\end{center}
\end{figure}

Let $\A=\{A_1,\ldots,A_n\}$ be a collection of $n$, possibly intersecting, 
obstacles $A_i$, completely contained in the region $R$, and let $B$ be the 
object moving relative to $\A$ under translations. See Figure~\ref{pr14dd}(a).
 (For notational convenience, we also denote $\A=\bigcup_{i=1}^n A_i$.)
%---------------------------------------------------------------------------
%----------------------------------------------------------------------------
%%---------------------------------------------------------------------------
%\begin{figure}[htb]
%\begin{center}
%\input{pres_14d.pstex_t}
%\caption{The translational Findspace and Findpath problems in case where 
%$\A=\{A_1,A_2,A_3\}$. (a) Here $FP_B(\A,R)$ is a connected region, $P_B$, and 
%$SP_B(\A)$ are the safe path, and the shortest safe path, of a moving object 
%$B$ relative to $\A$ from point $s$ to point $g$, respectively, and 
%$SP_B[FP_B(\A,R)]=SP_B(\A)$. Dashed lines show $\bd [CO_B(\A)]$, and 
%$\bd [CI_B(R)]$, respectively. (b) Illustration for the Problem IV. Here 
%$N_{IV}(B,\A,R)$ is a two-connected region and $\l_1=0$, $\l_2=|\l_R|$, and 
%$\l_3=1.5 \cdot\l_2$, respectively. Dashed curves show 
%$\bd [CO_B(\l_1,\l_2,\l_3,\A)]$, and $\bd [CI_B(\l_R,R)]$, respectively.}
%\label{pr14d}
%\end{center}
%\end{figure}
%%----------------------------------------------------------------------------

The {\it translational} Findspace problem \cite{LP} is to find all the 
possible translations $(B+p)\s R$, such that $(B+p)\cap\A=\es$. In this case 
$p$ is called the {\it free} position. (If $(B+p)\dot\cap \A$, then $p$ is 
called the {\it semi-free} position \cite{BKOS}.)
 
%To solve the Findspace problem in \cite{LPW} and \cite{LP} has 
%been  introduced the notion of {\it C-space obstacles}, that represents all 
%the {\it forbidden} positions of the object $B$ relative to the set of 
%obstacles $\A$. 

The C-space obstacle of $B$ relative to $\A$ is defined as 
$CO_B(\A)=\{p\mid (B+p)\cap\A\ne\es\}=\A\+\cB=\bigcup_{i=1}^n CO_B(A_i)
=\bigcup_{i=1}^n(A_i\+\cB)$ \cite{LP}. 
(The object $[CO_B(\A)]^c$ is called the {\it free C-space} of $B$ relative to 
$\A$.) The C-space {\it interior} of $B$ relative to the region $R$ is defined 
as $CI_B(R)=\{p\mid (B+p)\se R\}=[CO_B(R^c)]^c$; see Figure~\ref{pr14dd}(a). 
By the definition of the Minkowski difference, we get $CI_B(R)=R\-\cB$.

%The shortest safe path $SP_B(\A)$ of $B$ relative to $\A$ is the shortest 
%path in the {\it visibility graph} $VG_B[CO_B(\A)]$, obtained from the 
%C-space obstacle $CO_B(\A)$ by connecting all pair of its vertices (and $s$, 
%and $g$) that can ''see'' each other \cite{LPW}. See Figure~\ref{pr14d}.

In \cite{G} it is shown that the set of all the feasible positions of $B$ 
relative to $\A$ and $R$ can be represented as
$FP_B(\A,R)=(R\bsl\A)\- \cB=(R\-\cB)\bsl (\A \+\cB)=CI_B(R)\bsl CO_B(\A)$; 
see Figure~\ref{pr14dd}(a).
%Then $SP_B[FP_B(\A,R)]$ is the shortest path in the visibility graph 
%$VG_B[FP_B(\A,R)]$, obtained from the region $FP_B(\A,R)$. See 

Let us consider the Findspace problem in terms of the translational geometric 
situations (TGS), studied in Sections 7 and 8. The conditions $(B+p)\s R$ and 
$(B+p)\cap \A = \es$ can be formalized as $[\eta_1(B+p,R)\le 0]$ and 
$[\g_1(B+p,\A)>0]$, respectively. In \cite{Pk1} it is shown that 
$[\g_1(B+p,\A)>0]=\bigwedge_{i=1}^n[\g_1(B+p,A_i)>0]$. Then the Findspace 
problem can be reformulated as follows: Find the region $N(B,\A,R)$ 
corresponding to the TGS
$$\nu(B+p,\A,R)=\Big[\eta_1(B+p,R)\le 0\Big]\bigwedge
\bigg\{\bigwedge_{i=1}^n \Big[\g_1(B+p,A_i)>0\Big]\bigg\}.$$
Since $[\g_1(B+p,A_i)>0]=[\g_1(B+p,A_i)\le 0]^c$, we have 
$\bigwedge_{i=1}^n \big[\g_1(B+p,A_i)>0\big]=
\big\{\bigvee_{i=1}^n [\g_1(B+p,A_i)\le 0]\big\}^c$. Hence, the solving of the 
Findspace problem can be represented as follows:
$$N(B,\A,R)=H_1(0,K,B,R)\bigcap
\bigg\{\bigcap_{i=1}^n \Big[\G_1(0,K,B,A_i)\Big]^c\bigg\}$$
$$~~~~~~~~~~~~~
=H_1(0,K,B,R)\bigcap\Big[\bigcup_{i=1}^n \G_1(0,K,B,A_i)\Big]^c$$
$$~~~~~~~~~~~~~~~
=H_1(0,K,B,R)\bigcap\bigg[\G_1\Big(0,K,B,\bigcup_{i=1}^n A_i\Big)\bigg]^c$$
$$=H_1(0,K,B,R)\bsla \G_1(0,K,B,\A).$$
Clearly, $H_1(0,K,B,R)=CI_B(R)$ and $\G_1(0,K,B,\A)=CO_B(\A)$.

%For given TGS $\nu(B+p,\A,R)$ the safe path $P_B$ belongs to the region 
%$N(B,\A,R)$, and the shortest safe path $SP_B[N(B,\A,R)]$ of $B$ relative to 
%$N(B,\A,R)$ is the shortest path in the visibility graph $VG_B[N(B,\A,R)]$. 
%In case where $N(B,\A,R)$ is a disconnected region, the motion of $B$ among 
%the obstacle $\A$, for given TGS $\nu(B+p,\A,R)$, it is possible if and only 
%if the points $s$ and $g$ belong to the same connected component of 
%$N(B,\A,R)$.

We next consider more general Findspace problems.
% taking into account the $T$-distances $\o(B+p,A_i)$, between the moving 
%object $B$, and the obstacles %$A_i$, and the minimum distance 
%$\eta_1(B+p,R)$ between $B$ and the region $R$.

{\bf Problem III} Find the corresponding region $N_{III}(B,\A,R)$ of the TGS
$$\nu_{III}(B+p,\A,R)=\Big[ \eta_1(B+p,R)\le\l_R\le 0\Big]\bigwedge
\bigg\{\bigwedge_{i=1}^n \Big[\g_1(B+p,A_i)>\l_i\ge 0\Big]\bigg\}.$$
{\bf Solving the Problem III.} For given TGS $\nu_{III}(B+p,\A,R)$, the 
C-space obstacle depends on $\l_{\A}=\{\l_i\}_{i=1}^n$, corresponds to the TGS 
$\bigvee_{i=1}^n [\g_1(B+p,A_i)\le \l_i]$, and therefore it can be represented 
as $CO_B(\l_{\A},\A)=\bigcup_{i=1}^n \G_1(\l_i,K,B,A_i)$. The interior C-space 
corresponding to the TGS $[\eta_1(B+p,R)\le\l_R \le 0]$ is the  region 
$CI_B(\l_R,R)=H_1(\l_R,K,B,R)$. (See Sections 4, 5, and 
Figure~\ref{pr14dd}(b).) Thus, we get
$$N_{III}(B,\A,R)=H_1(\l_R,K,B,R)\bigcap
\bigg\{\bigcap_{i=1}^n \Big[\G_1(\l_i,K,B,A_i)\Big]^c\bigg\}$$
$$~~~~~~~~~~~~~~
=H_1(\l_R,K,B,R)\bsla\Big[\bigcup_{i=1}^n \G_1(\l_i,K,B,A_i)\Big].$$

\begin{figure}[htb]
\begin{center}
\input{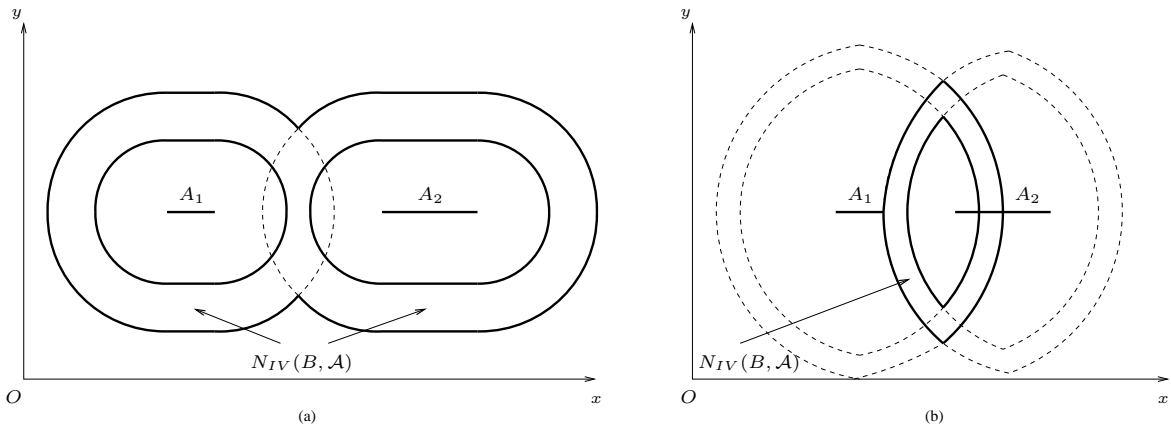}
\caption{Illustration for the Problem IV. Here $\A=\{A_1,A_2\}$, both $A_1$ 
and $A_2$ are line segments, $B=\{O\}$, $\l_1>\l_2$, and the region 
$N_{IV}(B,\A)$ is connected. Dashed curves show pieces of 
$\bd \G_{l_1(l_2)}(\l_{1(2)},B,A_{1(2)})$. Here (a) $l_{1(2)}=1$, and 
(b) $l_{1(2)}=2$.}
\label{pr14c}
\end{center}
\end{figure}

{\bf Problem IV} Find the solution $N_{IV}(B,\A)$ of the TGS
$$\nu_{IV}(B+p,\A)=\Big[\g_{l_1}(B+p,\A)\le \l_1\Big] \bigwedge
\Big[\g_{l_2}(B+p,\A)>\l_2\Big],\mbox{~where~}l_{1(2)}=1,2;~\l_{1(2)}\ge 0.$$
{\bf Solving the Problem IV.} The general solution is the region 
$$N_{IV}(B,\A)=\G_{l_1}(\l_1,K,B,\A)\bigcap \Big[\G_{l_2}(\l_2,K,B,\A)\Big]^c=
\G_{l_1}(\l_1,K,B,\A)\bsl\G_{l_2}(\l_2,K,B,\A).$$
By observations of \cite{Pk1} and of Section 8, in case where $l_{1(2)}=1$ and 
$\l_{1(2)}\ge 0$, we have
$$\nu_{IV}(B+p,\A)=
\bigg\{\bigvee_{i=1}^n \Big[\g_1(B+p,A_i)\le \l_1\Big]\bigg\} \bigwedge
\bigg\{\bigwedge_{i=1}^n \Big[\g_1(B+p,A_i)>\l_2\Big]\bigg\};$$
$$N_{IV}(B,\A)=\Big[\bigcup_{i=1}^n \G_1(\l_1,K,B,A_i)\Big]
\bsla \Big[\bigcup_{i=1}^n \G_1(\l_2,K,B,A_i)\Big],$$
whereas, for $l_{1(2)}=2$ and 
$\l_{1(2)}\ge\min_{1\le i\le n}\{R_{A_i\+\cB}\}$, we get
$$\nu_{IV}(B+p,\A)=
\bigg\{\bigwedge_{i=1}^n \Big[\g_2(B+p,A_i)\le \l_1\Big]\bigg\}\bigwedge
\bigg\{\bigvee_{i=1}^n \Big[\g_2(B+p,A_i)>\l_2\Big]\bigg\};$$
$$N_{IV}(B,\A)=\Big[\bigcap_{i=1}^n \G_2(\l_1,K,B,A_i)\Big]
\bsla\Big[\bigcap_{i=1}^n \G_2(\l_2,K,B,A_i)\Big].$$
See Figure~\ref{pr14c}. Note that $\nu_{IV}(B+p,\A)=0$ and $N_{IV}(B,\A)=\es$, 
for $\l_1\le\l_2$.
%Recall that $\G_1(\l,K,B,\A)\sp \bigcup_{i=1}^n \G_1(\l,K,B,A_i)$, 
%for $\l<0$. Therefore in case where $l_{1,2}=1$, and $\l_{1,2}<0$, the 
%solution $N_V(B,\A)$ can be represented as above only in special cases, e.g., 
%for pairwise disjoint objects $\{A_i\+ \cB\}_{i=1}^n $.
%---------------------------------------------------------------------------
%----------------------------------------------------------------------------

Finally, we consider the Findspace problem concerning the {\it covering} of 
objects.
%the obstacles $A_i$ by the moving object $B$ with constraints on distances 
%$\dl_{1,2}(B+p,A_i)$ between $B$ and $A_i$.

{\bf Problem V} Find the region $N_V(B+p,\A)$ of all the possible 
coverings of object $\A$ by object $B+p$, for given TGS
$$\nu_V(B+p,\A)=\bigg\{\bigvee_{i=1}^n
\Big[\dl_2(B+p,A_i)\le\e_i\Big]\bigg\}\bigwedge
\bigg\{\bigwedge_{i=1}^n
\Big[\l_i^-\le\dl_1(B+p,A_i)\le\l_i^+\le 0\Big]\bigg\}.$$
The solution is
$$N_V(B+p,\A)=\Big[\bigcup_{i=1}^n \Dl_2(\e_i,K,B,A_i)\Big]\bigcap
\bigg\{\bigcap_{i=1}^n
\Big[\Dl_1(\l_i^+,K,B,A_i)\bsla\i\Dl_1(\l_i^-,K,B,A_i)\Big]\bigg\}.$$
For more simple constraint 
$\nu_V(B+p,\A)=\bigwedge_{i=1}^n [\dl_1(B+p,A_i)\le\l_i\le 0]$, 
we get $N_V(B,\A)=\bigcap_{i=1}^n \Dl_1(\l_i,K,B,A_i)$; see 
Figure~\ref{pr14ee}.
%---------------------------------------------------------------------------
\begin{figure}[htb]
\begin{center}
\input{pres_14ee.pstex_t}
\caption{Illustration for the Problem V. (a) An objects $\A=\{A_1,A_2,A_3\}$ 
and $B$. (b) Here the region $N_V(B,\A)$ is siply connected, $\A\s(B+p)$, 
for $p\in N_V(B,\A)$; $\l_1=0$, $\l_2<0$, and $\l_3=\l_2$, respectively. 
Dashed lines show $\bd \Dl_1(\l_{1-3},B,A)$.}
\label{pr14ee}
\end{center}
\end{figure}
%----------------------------------------------------------------------------

\subsubsection*{9.2 Placement of geometric objects}

An approches commonly used for solving the placement problems are the 
{\it sequential-single} method, suggested in \cite{St} and \cite{SY}, and the 
{\it multiple} placement of several objects, suggested in \cite{AB}, 
\cite{DM}, and \cite{Ml1}.

The sequential-single placement consists in the sequential placement of the 
objects of $\A$ with respect to the {\it container} $A_0$ in a fixed order, 
say $A_1,A_2,\ldots,A_n$, according to the special objective function. For 
instance, the valid position $p_i$ of $A_i$ must be a point with extremal (or 
specified) values of coordinates. (In case of planar problems $p_i=(x_i,y_i)$ 
can have, for example, a minimal, maximal, or specified $x_i$ and/or $y_i$.) 

The multiple (or simultaneous) placement, as follows from its name, is 
independent on the order of placement of the objects of $\A$ relative to 
$A_0$, and provides the placement of each object of $\A$, say $A_j$, taking 
into account the possibility of the placement of all another objects 
$\{A_i\}$, where $1\le i\le n$, and $i\ne j$. The goal is to find the set 
$P_{0j}$ of all the feasible positions of $A_j$, for $j=1,\ldots,n$, with 
respect to $A_0$.

The generalized sequential-single and multiple placement problems (i.e., the 
problems with more complex constraints on the relative position of objects 
than in \cite{Ml1} and \cite{SY}) and their solutions have been studied in 
\cite{Pk1}.

\subsubsection*{9.3 Application to the other types of geometric situations}

The work in \cite{Pk3} has studied the following types of geometric situations:

The {\it translational} geometric situation {\it in direction $u$} describes 
constraints on translational distances {\it in a given direction} between 
geometric objects. The minimum and maximum distances 
%See Figure~\ref{pr17}.The minimum and maximum distances 
$\g_{1,2}(B,A,u)$ taking into account the {\it outer} position of $B$ relative 
to $A$ in direction $u$ have been introduced in \cite{KS} and \cite{P}. The 
minimum and maximum distances $\eta_{1,2}(B,A,u)$ taking into account the 
{\it inner} position of $B$ relative to $A$ have been proposed in \cite{Pk} 
and \cite{Pk3}. The parametric families of objects  corresponding to the 
distances in a given direction are obtained by using the {\it partial} 
vector operations on objects, which are generalizations of the Minkowski 
operations. See \cite{Pk3} and \cite{R} for more detailes.
%---------------------------------------------------------------------------
%\begin{figure}[htb]
%\begin{center}
%\input{pres_17.pstex_t}
%\caption{The translational distances in direction $u$ between geometric 
%objects $A$ and $B$. The minimum and maximum distances $\g_{1,2}(B,A,u)$ (a), 
%and $\eta_{1,2}(B,A,u)$ (b), taking into account the outer, and the inner 
%position of $B$ relative to $A$, respectively.}
%\label{pr17}
%\end{center}
%\end{figure}
%----------------------------------------------------------------------------

The {\it rotational} geometric situation describes constraints on minimum 
and maximum {\it rotational} distances between geometric objects. Denote by 
$A^*$ the image of $A$ in polar coordinates, i.e., 
$A^*=\{(r,\t) \mid (r\cos\t,r\sin\t) \in A\}$. Then a copy $A^{\phi}$ of $A$ 
rotated by an angle $\phi$ around the origin $O$ corresponds to a copy 
$A^*+\{(0,\phi)\}$ of $A^*$ translated by a point $(0,\phi)$ in polar 
coordinates. Therefore one can use the {\it partial} vector operations in 
polar coordinates for modeling the relative position of geometric objects 
under rotations. The rotational distances and the translational distances in a 
given direction have used in \cite{Pk3} to formalize the constraints on the 
relative position of links in problems of modeling of mechanism's motion.
%---------------------------------------------------------------------------
%\begin{figure}[htb]
%\begin{center}
%\input{pres_18.pstex_t}
%\caption{The rotational distances between geometric objects $A$ and $B$. 
%The minimum and maximum distances $\g^*_{1,2}(B,A)$ (a), and $\eta^*_1(B,A)$ 
%(b), taking into account the outer, and the inner position of $B$ relative to 
%$A$, respectively.}
%\label{pr18}
%\end{center}
%\end{figure}
%----------------------------------------------------------------------------

The {\it dynamic} geometric situation is defined, for moving objects, by 
representing an objects as a four-dimensional sets in the space-time $G^4$; 
see \cite{C}, \cite{Pk}, and \cite{Pk3}. (Note that the distances between 
objects ``in the space'' and ``in the time'' are incomparable \cite{Y}.) 

%Let $A^*,B^*$ be the images of $A,B$ in $G^4$:
%$$A^*=\cup_{t\in [0,1]}[A(t)],~~~ B^*=\cup_{t\in [0,1]}[B(t)].$$
Let $A^*=\bigcup_{t\in [0,1]}[A(t)]$, $B^*=\bigcup_{t\in [0,1]}[B(t)]$ be the 
image of $A,B$ in $G^4$. We denote by $\mathop \+^t$ the partial addition 
``by the time'', and by $\cB^*$ the reflection of $B^*$ with respect to the 
origin $O$ in $R^3$. Let next $\l K^*$ be the cylinder in $G^4$ with a basis 
$\l K$. Then the parametric family of objects 
$$\G_1(\l,K,B^*,A^*)=\bigcup_{t\in [0,1]}[A(t)\+ \cB (t)\+\l K]=
A^* \mathop \+^t{\cB}^*\mathop \+^t\l K^*$$
corresponds to the distance 
$\g_1(B^*,A^*)=\inf_{t\in [0,1]}\{\g_1[B(t),A(t)]\}$ between $A^*$ and $B^*$. 
%(Note that can be considered the special case where $\l=\l(t)$.) 

%The dynamic geometric situations are obtained analogously to the 
%translational geometric situations. Algorithms for solving the dynamic 
%geometric situations are given in \cite{Pk} and \cite{Pk3}.

It is clear that the suggested technique for solving the translational spatial 
planning problems can also be applied to geometric problems with the 
considered  types of constaints on the relative position of objects.

\subsection*{10 Computational issues}

From the observations of Sections 8 and 9 it follows that the C-space map of 
spatial planning problem, is the region obtained by standard and/or 
regularized Boolean operations, and by Minkowski operations on regular and/or 
non-regular objects. In general, the C-space map is a {\it non-regular} 
geometric object of various topology with {\it non-manifold} boundary; see 
Remark 5 of Section 8. (Recall that the non-regular object may have external 
dangling faces/edges and/or isolated points, and/or internal entities such as 
cracks and/or isolated points \cite{Rq}.) Therefore for implementation of 
spatial planning we need the methods for representation and manipulation of 
such an objects. In this section we briefly consider the computer 
representations of geometric objects that are suitable for solving the spatial 
planning problems, and the strategies for computing the C-space maps.

\paragraph{Representaions of geometric objects.}The two representation schemes 
that are most widely used in solid modeling and computer graphics are boundary 
representation (BRep) and constructive solid geometry (CSG) \cite{Rq}. Let 
$A$ be a point set of $R^n$ ($n=2,3$). CSG($A$) is a Boolean composition of 
algebraic halfspaces using regularized set operations. BRep($A$) is a 
collection of closed faces/edges. The problems of CSG to BRep conversion and 
of BRep to CSG conversion have been studied in \cite{Rq}, \cite{ShV1}, and 
\cite{ShV2}.

The third type of representation scheme suitable for our purpose is the linear 
ray representation (LRRep) \cite{MMZ}, \cite{Pk}, denoted by LRR($A$). (In 
\cite{Pk} it is called the linear raster representation.) LRR($A$) is an 
approximation of an object $A$ by a set of parallel segments belonging to a 
grid $L$ of parallel lines, i.e., LRR($A$)$=A\cap L$. Conversions between BRep 
and LRRep, and between CSG and LRRep have been detally studied in \cite{MV}.

The constructive non-regularized geometry (CNRG) methodology for 
representation and manipulation of non-homogeneous (i.e., made of several 
materials with different properties), non-closed point sets with internal 
structures and incomplete boundaries have been suggested in \cite{RR2}. The 
work in \cite{GCP} has proposed an approach for representation of geometric 
objects with non-manifold boundary.

The boundary representation of non-regular geometric objects with non-manifold 
boundary using the techniques of \cite{GCP} and \cite{RR2} have been studied 
in \cite{Pk1}. It takes into account both the geometry and the topology of 
objects. The work in \cite{Pk1} have also considered the topological 
operations (complement, interior, closure, and regularization) on a single 
non-regular object.

\paragraph{Boolean operations.} Algorithms and implementation for computing 
the regularized set operations on polyhedral objects have been proposed in 
\cite{RV}. Recall that the standard union $A\cup B$ of two $r$-sets $A$ and 
$B$ always results in an $r$-set, whereas the standard intersection $A\cap B$ 
and the standard set difference $A\bsl B$ need not be regular: the set 
$A\cap B$ may have dangling edges, e.g., in case where $A$ contacts with $B$ 
along the portion of its boundary \cite{Rq}; the set $A\bsl B$ may be 
partially open, e.g., in case where $\i A\cap \i B\ne\es$ \cite{ORK}, 
\cite{RV}. Algorithms for computing the set operations on non-manifold 
boundary representation objects have been proposed in \cite{GCP} and 
\cite{RO}. We assume below some familiarity with theory and algorithms of 
\cite{GCP} and \cite{RV}.

As mentioned in \cite{RV}, the algorithms will work with curved objects, and 
they are insensitive to whether a solid's boundary is or is not a 
two-manifold, and is or is not connected. Hence, the algorithms of \cite{RV} 
can be modified to compute the standard Boolean operations on non-regular 
objects of various topology with non-manifold boundary. 

Let $S=A\otimes^* B$, where $\otimes$ denotes one of the standard set 
operations. The main utilities used in algorithms of \cite{RV} to compute the 
boundary of $S$ are the set membership classification (SMC) \cite{T}, and the 
combining classifications, defined by means of the regularized set operations. 
See \cite{RV} and \cite{T} for details.

In \cite{Pk1} the algorithms of \cite{RV} have been modified for computing the 
standard Boolean set operations on non-regular (possibly unbounded) geometric 
objects of various topology, for BReps. To define and to combine the 
classifications the work in \cite{Pk1} have used the standard, but not a 
regularized set operations.

\paragraph{Minkowski operations.}  Many various algorithms to compute the 
Minkowski operations have been proposed. Detailed surveys of previous work on 
computing the Minkowski operations can be found in \cite{FoH}, \cite{Hl}, 
\cite{LKE}, \cite{LM}, \cite{ORK}, \cite{VM}, and \cite{WB}. Algorithms for 
computing the Minkowski sums and the Minkowski differences in two and three 
dimensions are given, e.g., in \cite{AB}, \cite{BK}, \cite{G}, \cite{G1}, 
\cite{GRS}, \cite{LKE} -- \cite{LP}, \cite{Ml1}, \cite{P}, and \cite{VM}, for 
BReps, and in \cite{MMZ} and \cite{Pk}, for LRReps. Note that the referenced 
algorithms perform computing the Minkowski operations on regular objects. They 
can generate the manifold boundaries and are not applicable to the cases where 
the boundary of the resulting object is non-manifold; see \cite{VM} for 
details. 

In \cite{FH} and \cite{Hl} have been presented an algorithms for {\it robust} 
and {\it efficient} construction of planar Minkowski sums for polygons using 
exact rational arithmetic. In contrast with most existing techniques the 
algorithms of \cite{FH}, \cite{Hl} directly handle the degenerate 
configurations, arising in the boundary of the Minkowski sum, such as internal 
isolated points and/or coinciding edges. 
%(See Figure 3 of \cite{FH} and Figure 1 of \cite{Hl}, respectively.) 
In other words, these algorithms compute the outer envelope of $A$ and $B$, 
i.e., the boundary of the open set $\abi$ (see subsection 2.1). The recent 
works in \cite{W1} and \cite{W} have presented an algorithms for exact and 
efficient construction of Minkowski sums of polygons, and for exact and 
approximate construction of offset polygons, respectively, that handle the 
degenerate configurations also. Hence, the algorithms of \cite{FH}, \cite{Hl}, 
\cite{W1}, and \cite{W} allow to construct the parametric families of 
polygonal objects used for computing the C-space maps. 

The algorithms of \cite{FH} and \cite{Hl} are based on convex decomposition of 
polygons. However, as mentioned in \cite{BK}, not all curved objects permit 
convex decomposition, e.g., an object with an inward concave edge. Therefore 
to handle the curved objects more suitable are the methods that deal with the 
geometric objects directly. 

In \cite{Pk1} the algorithms of \cite{BK} have been modified for computing the 
Minkowski operations on non-regular objects of various topology, for BReps, 
using the techniques of \cite{GCP} and \cite{RR2}.

\paragraph{Distances between geometric objects.} Many algorithms for computing 
the distances between geometric objects have been developed (see \cite{LinM}). 
Detailed surveys of previous work on computing the MTD between regular objects 
can be found in \cite{DHKS}, \cite{FoH}, \cite{GJK}, \cite{JTT}, \cite{LinM}, 
and \cite{OG}. Algorithms for computing the distances concerning the outer 
relative position of objects in two and three dimensions (see subsection 2.2) 
are given, e.g., in \cite{CC}, \cite{DHKS}, \cite{FoH} -- \cite{GRS}, 
\cite{OG}, \cite{P}, and \cite{Sr}, for BReps, and in \cite{Pk} and 
\cite{Pk3}, for LRReps.  The algorithms of \cite{FH}, \cite{Hl}, and \cite{W1} 
for robust construction of planar Minkowski sums can be used for computing the 
MTD between non-regular polygonal objects. 

Algorithms for computing the distances concerning the containment of objects 
in two and three dimensions (see subsection 5.1) are given in \cite{Pk} and 
\cite{Pk3}, for LRReps. Note that for this goal can also be used algorithms 
for computing the Minkowski difference, for BReps; see, e.g., \cite{G} and 
\cite{G1}. Algorithms for computing the distances concerning the covering of 
objects (see subsection 5.2) and algorithms for computing the distances 
concerning the containment of objects are similar.

Algorithms for computing the translational distances in a given direction in 
two and three dimensions are given in \cite{DHKS}, \cite{FoH}, \cite{KS}, and 
\cite{Sr}, for BReps, and in \cite{Pk} and \cite{Pk3}, for LRReps, 
respectively. The algorithms of \cite{Pk} and \cite{Pk3} are based on the 
partial vector operations (see subsection 9.1). Algorithms for computing the 
rotational distances and the partial vector operations in polar coordinates 
are given in \cite{Pk3}.

Thus, using the algorithms for computing the various types of distances 
between geometric objects provide solving the generalized distance query 
problem, as defined in subsection 1.3.

\subsection*{Acknowledgments}
The author would like to thank Prof. Micha Sharir for his support, assistance, 
and advice on this work; for his useful suggestions in the content and 
presentation of this paper, and for helpful comments.

%\newpage

\end{document}